\documentclass[12pt]{article}
\usepackage{axodraw,epsf}

\def\slashchar#1{\setbox0=\hbox{$#1$}           
   \dimen0=\wd0                                 
   \setbox1=\hbox{/} \dimen1=\wd1               
   \ifdim\dimen0>\dimen1                        
      \rlap{\hbox to \dimen0{\hfil/\hfil}}      
      #1                                        
   \else                                        
      \rlap{\hbox to \dimen1{\hfil$#1$\hfil}}   
      /                                         
   \fi}                                         %

\def\stilde{\widetilde}
\def\conj{{\rm c.c.}}
\newcommand{\newc}{\newcommand}
\newc{\gsim}{\lower.7ex\hbox{$\;\stackrel{\textstyle>}{\sim}\;$}}
\newc{\lsim}{\lower.7ex\hbox{$\;\stackrel{\textstyle<}{\sim}\;$}}
\newc{\oneloop}{h}
\newc{\Nindex}{n}
\newc{\Iindex}{n}
\newc{\Jindex}{p}
\newc{\Kindex}{q}
\newc{\Lindex}{r}
\newc{\Nc}{N_{c}}
\newc{\CG}{C_G}
\newc{\Ni}{\tilde{N}_i}
\newc{\Nj}{\tilde{N}_j}
\newc{\Ci}{\widetilde{C}_i}
\newc{\Cj}{\widetilde{C}_j}
\newc{\gp}{g'}
\newc{\stopi}{\stilde t_i}
\newc{\sboti}{\stilde b_i}
\newc{\staui}{\stilde \tau_i}
\newc{\stopj}{\stilde t_j}
\newc{\sbotj}{\stilde b_j}
\newc{\stauj}{\stilde \tau_j}
\newc{\stopI}{\stilde t_1}
\newc{\stopII}{\stilde t_2}
\newc{\sbotI}{\stilde b_1}
\newc{\sbotII}{\stilde b_2}
\newc{\stauI}{\stilde \tau_1}
\newc{\stauII}{\stilde \tau_2}
\newc{\sstop}{s_{t}}
\newc{\cstop}{c_{t}}
\newc{\ssbot}{s_{b}}
\newc{\csbot}{c_{b}}
\newc{\sstau}{s_{\tau}}
\newc{\cstau}{c_{\tau}}
\newc{\Sstop}{s_{2t}}
\newc{\Cstop}{c_{2t}}
\newc{\Ssbot}{s_{2b}}
\newc{\Csbot}{c_{2b}}
\newc{\Sstau}{s_{2\tau}}
\newc{\Cstau}{c_{2\tau}}
\newc{\salpha}{s_\alpha}
\newc{\calpha}{c_\alpha}
\newc{\Calpha}{c_{2\alpha}}
\newc{\Salpha}{s_{2\alpha}}
\newc{\sbetapm}{s_{\beta_\pm}}
\newc{\cbetapm}{c_{\beta_\pm}}
\newc{\Sbetapm}{s_{2 \beta_\pm}}
\newc{\Cbetapm}{c_{2 \beta_\pm}}
\newc{\sbetaO}{s_{\beta_0}}
\newc{\cbetaO}{c_{\beta_0}}
\newc{\SbetaO}{s_{2 \beta_0}}
\newc{\CbetaO}{c_{2 \beta_0}}
\newc{\vu}{v_u}
\newc{\vd}{v_d}
\newc{\seL}{\stilde e_L}
\newc{\smuL}{\stilde \mu_L}
\newc{\seR}{\stilde e_R}
\newc{\smuR}{\stilde \mu_R}
\newc{\suL}{\stilde u_L}
\newc{\sdL}{\stilde d_L}
\newc{\suR}{\stilde u_R}
\newc{\sdR}{\stilde d_R}
\newc{\scL}{\stilde c_L}
\newc{\ssL}{\stilde s_L}
\newc{\scR}{\stilde c_R}
\newc{\ssR}{\stilde s_R}
\newc{\snue}{\stilde \nu_e}
\newc{\snumu}{\stilde \nu_\mu}
\newc{\snutau}{\stilde \nu_\tau}
\newc{\Gpm}{G^\pm}
\newc{\Hpm}{H^\pm}
\newc{\FFbS}{\overline{FF}S}
\newc{\FFbV}{\overline{FF}V}
\newc{\FSS}{F_{SS}}
\newc{\FSSS}{F_{SSS}}
\newc{\FFFS}{F_{FFS}}
\newc{\FFFbS}{F_{\overline{FF}S}}
\newc{\FSSV}{F_{SSV}}
\newc{\FVS}{F_{VS}}
\newc{\FVVS}{F_{VVS}}
\newc{\FFFV}{F_{FFV}}
\newc{\FFFbV}{F_{\overline{FF}V}}
\newc{\Fgauge}{F_{\rm gauge}}
\newc{\DRbarprime}{$\overline{\rm DR}'$ }
\newc{\DRbar}{$\overline{\rm DR}$ }
\newc{\MSbar}{$\overline{\rm MS}$ }
\newc{\Yu}{{\bf Y}_u}
\newc{\Yd}{{\bf Y}_d}
\newc{\Ye}{{\bf Y}_e}
\newc{\Au}{{\bf a}_u}
\newc{\Ad}{{\bf a}_d}
\newc{\Ae}{{\bf a}_e}
\newc{\bm}{{\bf m}}
\def\beq{\begin{eqnarray}}
\def\eeq{\end{eqnarray}}
\def\bea{\begin{eqnarray}}
\def\eea{\end{eqnarray}}
\newcommand{\captionfonts}{\small}

\makeatletter  
\long\def\@makecaption#1#2{%
  \vskip\abovecaptionskip
  \sbox\@tempboxa{{\captionfonts #1: #2}}%
  \ifdim \wd\@tempboxa >\hsize
    {\captionfonts #1: #2\par}
  \else
    \hbox to\hsize{\hfil\box\@tempboxa\hfil}%
  \fi
  \vskip\belowcaptionskip}
\makeatother   

\begin{document}
\begin{titlepage}
\begin{flushright}
hep-ph/0206136 \\
Fermilab-Pub-02/118-T
\end{flushright}
\vspace{0.3in}

\begin{center}
{\large\bf
Two-loop effective potential for the minimal supersymmetric 
standard model}
\end{center}
\vspace{.15in}

\begin{center}
{\sc Stephen P.~Martin}

\vspace{.1in}
{\it Department of Physics, Northern Illinois University, DeKalb IL 60115
{$\rm and$}\\
}{\it Fermi National Accelerator Laboratory,
P.O. Box 500, Batavia IL 60510 \\} 
\end{center}

\vspace{0.8in}

\begin{abstract}\noindent I compute the complete two-loop effective
potential for the minimal supersymmetric standard model in the Landau
gauge. This enables an accurate determination of the minimization
conditions for the vacuum expectation values of the Higgs fields. Checks
on the result follow from supersymmetric limits and from
renormalization-scale invariance. The renormalization group equations for
the field-independent vacuum energy and the vacuum expectation values are
also presented. I provide numerical examples showing the improved accuracy
and scale dependence obtained with the full two-loop effective potential.

\end{abstract}
\end{titlepage}
\baselineskip=19pt

\setcounter{page}{2}
\setcounter{figure}{0}
\setcounter{table}{0}

\tableofcontents

\section{Introduction}
\label{sec:intro}
\setcounter{equation}{0}
\setcounter{footnote}{1}

The mechanism of electroweak symmetry breaking will be the principal focus
of experimental investigations at the high-energy frontier for the next
decade.  Supersymmetry provides a highly predictive mechanism for
addressing the hierarchy problem associated with the electroweak symmetry
breaking scale.  If supersymmetry is correct, then interpretations of
future experimental data will rely on precise theoretical calculations in
candidate models of supersymmetry breaking. The effective potential
\cite{Coleman:1973jx}-\cite{Sher:1989mj} approach allows the computation
of the vacuum expectation values (VEVs) of Higgs fields in terms of the
underlying Lagrangian parameters of a given theory.

The scalar
potential of the Minimal Supersymmetric Standard Model (MSSM;  
for reviews, see \cite{Haber:1984rc}-\cite{primer}) is notoriously 
sensitive to radiative corrections. At tree
level, the Higgs field quartic couplings are proportional to a sum of
squares of electroweak gauge couplings and are therefore known not to be
very large. Furthermore, they actually vanish along a $D$-flat direction
in field space. These facts ensures the existence of at least one light
Higgs scalar boson, corresponding to a shallow direction in the effective
potential for the Higgs vacuum expectation values.  The same facts also
imply that the minimization conditions for the scalar potential depend
very significantly on radiative corrections.

Previous results for the effective potential in the MSSM have included the
full one-loop contributions \cite{MSSMonelooppot} and partial two-loop
corrections \cite{Hempfling:1994qq}-\cite{Espinosa:2000df} including the
effects of 
the QCD coupling and the top and bottom Yukawa couplings.
Including these contributions mitigates the scale-dependence of
the tree-level effective potential. However, I find that there is still a 
significant
scale dependence and overall error compared to the uncertainties in 
theoretical quantities
that may eventually be obtained at future experiments, especially at a
linear $e^+e^-$ collider.
In this paper, I will present the result for the full two-loop effective
potential of the MSSM, in the Landau gauge and in the \DRbarprime
\cite{Jack:1994rk} regularization and renormalization scheme. This is an
application of the results given in ref.~\cite{general} for a general
field theory. The
\DRbarprime scheme is the variant of the \DRbar scheme \cite{DRED} in 
which
the effects of unphysical epsilon-scalars masses are removed by parameter
redefinitions \cite{Jack:1994rk,general}. 

The two-loop effective potential
for a general renormalizable theory can be written as
\beq
V_{\rm eff} = V^{(0)} + {1\over 16 \pi^2} V^{(1)} +
{1\over (16 \pi^2)^2} V^{(2)} .
\eeq
Here $V^{(0)}$ is the tree-level contribution. In the \DRbarprime 
scheme, the one-loop contribution is
\beq
V^{(1)} = \sum_n (-1)^{2 s_n} (2 s_n+1) h(m_n^2),
\eeq
where $s_n=0,1/2,1$ for real scalars, two-component fermions, and
vector fields, respectively, with field-dependent tree-level squared 
masses $m_n^2$. 
The one-loop function is
\beq
h(x) = {x^2 \over 4} \left [{\rm ln}(x/Q^2) - 3/2 \right ],
\label{defineh}
\eeq
where $Q$ is the renormalization scale. 
The two-loop contribution always has the form
\beq
V^{(2)} = \sum_{n,p} \lambda^{nnpp} F_{np}(m_n^2, m_p^2) +
\sum_{n,p,q} |\lambda^{npq}|^2 F_{npq}(m_n^2,m_p^2,m_q^2),
\eeq
where $\lambda^{npqr}$ and $\lambda^{npq}$ are tree-level field-dependent 
four- and 
three-particle couplings, and $F_{np}$ and $F_{npq}$ are $Q$-dependent
functions of the $m_n^2$, depending on the particle types. The
two-loop functions can be evaluated analytically using various
methods developed in \cite{Kotikov}-\cite{Caffo:1998du}.
In general, one can write the results in terms of 10
basis functions, corresponding to 
the one-particle-irreducible connected vacuum graphs shown in Figure 1.
These functions were given explicitly in ref.~\cite{general}.
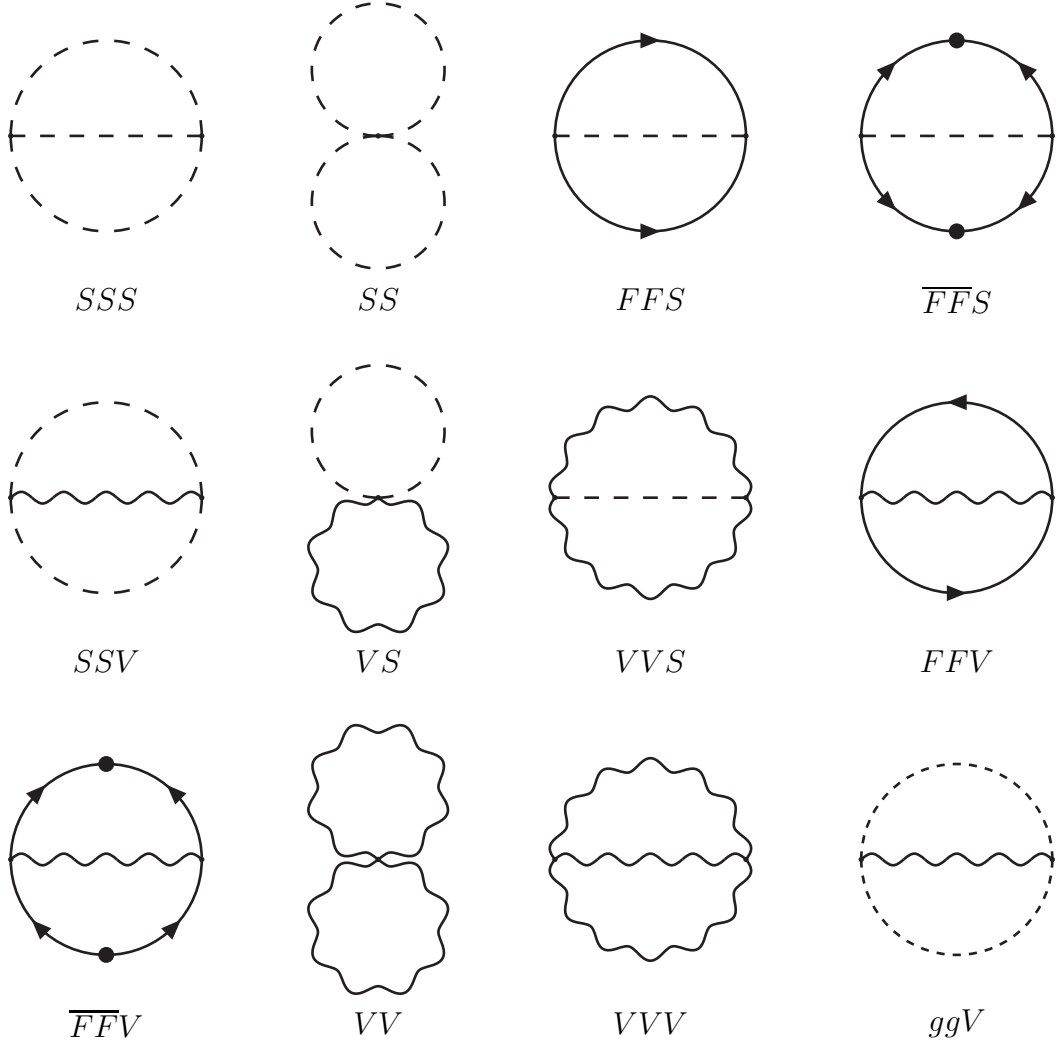
\begin{figure}[tbp]
\SetWidth{1}
\begin{picture}(112,136)(-56,-70)
\DashCArc(0,0)(36,0,180){6}
\DashCArc(0,0)(36,180,360){6}
\DashLine(-36,0)(36,0){6}
\Vertex(-36,0){1}
\Vertex(36,0){1}
\Text(0,-62)[]{$SSS$}
\end{picture}
\begin{picture}(86,136)(-43,-70)
\DashCArc(0,25)(25,-90,270){6}
\DashCArc(0,-25)(25,90,450){6}
\Vertex(0,0){1}
\Text(0,-62)[]{$SS$}
\end{picture}
\begin{picture}(112,136)(-56,-70)
\ArrowArcn(0,0)(36,180,0)
\ArrowArc(0,0)(36,180,360)
\DashLine(-36,0)(36,0){6}
\Vertex(-36,0){1}
\Vertex(36,0){1}
\Text(0,-62)[]{$FFS$}
\end{picture}
\begin{picture}(112,136)(-56,-70)
\ArrowArcn(0,0)(36,180,90)
\ArrowArc(0,0)(36,-180,-90)
\ArrowArc(0,0)(36,0,90)
\ArrowArcn(0,0)(36,0,-90)
\DashLine(-36,0)(36,0){6}
\Vertex(-36,0){1}
\Vertex(36,0){1}
\Vertex(0,-36){3}
\Vertex(0,36){3}
\Text(0,-62)[]{$\overline{FF}S$}
\end{picture}

\noindent
\begin{picture}(112,136)(-56,-70)
\DashCArc(0,0)(36,0,180){6}
\DashCArc(0,0)(36,180,360){6}
\Photon(-36,0)(36,0){2.2}{4.5}
\Vertex(-36,0){1}
\Vertex(36,0){1}
\Text(0,-62)[]{$SSV$}
\end{picture}
\begin{picture}(86,136)(-43,-70)
\DashCArc(0,25)(25,-90,270){6}
\PhotonArc(0,-25)(25,90,450){-2.2}{8.5}
\Vertex(0,0){1}
\Text(0,-62)[]{$VS$}
\end{picture}
\begin{picture}(112,136)(-56,-70)
\PhotonArc(0,0)(36,0,180){2.2}{6.5}
\PhotonArc(0,0)(36,180,360){2.2}{6.5}
\DashLine(-36,0)(36,0){6}
\Vertex(-36,0){1}
\Vertex(36,0){1}
\Text(0,-62)[]{$VVS$}
\end{picture}
\begin{picture}(112,136)(-56,-70)
\ArrowArc(0,0)(36,0,180)
\ArrowArc(0,0)(36,180,360)
\Photon(-36,0)(36,0){2.2}{4.5}
\Vertex(-36,0){1}
\Vertex(36,0){1}
\Text(0,-62)[]{$FFV$}
\end{picture}

\noindent
\begin{picture}(112,136)(-56,-70)
\ArrowArcn(0,0)(36,180,90)
\ArrowArc(0,0)(36,0,90)
\ArrowArcn(0,0)(36,270,180)
\ArrowArc(0,0)(36,-90,0)
\Photon(-36,0)(36,0){2.2}{4.5}
\Vertex(-36,0){1}
\Vertex(36,0){1}
\Vertex(0,-36){3}
\Vertex(0,36){3}
\Text(0,-62)[]{$\overline{FF}V$}
\end{picture}
\begin{picture}(86,136)(-43,-70)
\PhotonArc(0,25)(25,-90,270){-2.2}{8.5}
\PhotonArc(0,-25)(25,90,450){-2.2}{8.5}
\Vertex(0,0){1}
\Text(0,-62)[]{$VV$}
\end{picture}
\begin{picture}(112,136)(-56,-70)
\PhotonArc(0,0)(36,0,180){2.2}{6.5}
\PhotonArc(0,0)(36,180,360){2.2}{6.5}
\Photon(-36,0)(36,0){2.2}{4.5}
\Vertex(-36,0){1}
\Vertex(36,0){1}
\Text(0,-62)[]{$VVV$}
\end{picture}
\begin{picture}(112,136)(-56,-70)
\DashCArc(0,0)(36,0,180){3}
\DashCArc(0,0)(36,180,360){3}
\Photon(-36,0)(36,0){2.2}{4.5}
\Vertex(-36,0){1}
\Vertex(36,0){1}
\Text(0,-62)[]{$ggV$}
\end{picture}
\caption{The one-particle-irreducible connected Feynman diagrams
contributing to the two-loop effective potential. Dashed lines
denote real scalars, solid lines denote Weyl fermions carrying
helicity along the arrow direction, wavy lines are for vector
bosons, and dotted lines are for ghosts. The large dots between
opposing arrows on the
fermion lines in the $\FFbS$ and $\FFbV$
diagrams denote mass insertions. The $\FFbS$ diagram
is accompanied by its complex conjugate (the same diagram with all arrows
reversed). The effects of the VV, VVV, and ggV diagrams can always be
combined into a ``gauge" contribution. The loop integral functions 
associated with these Feynman diagrams are given explicitly in 
ref.~\cite{general}.} \label{fig:feynmandiagrams}
\end{figure}

In the MSSM, there are two Higgs VEVs, $v_u$ and $v_d$. The evaluation of
the two-loop effective potential as a function of the VEVs reduces to
determination of the relevant tree-level field-dependent couplings and
masses. I will do this in the approximation of no Yukawa or
flavor-violating couplings for the first two families of (s)quarks and
(s)leptons. However, all remaining complex phases which cannot be rotated 
away
are maintained. The resulting expressions presented here are complicated,
but suitable for direct evaluation by computer programs.

The rest of this paper is organized as follows. In section
\ref{sec:setup}, I describe necessary conventions and define some
coefficients used in the calculation. Section \ref{sec:potential} contains
the expressions for the effective potential in the MSSM up to two-loop
order. 
Section \ref{sec:susylimits} discusses 
supersymmetric limits of the effective potential. 
In section \ref{sec:scale}, I present two-loop results for the
Higgs scalar anomalous dimensions and the beta function of the vacuum
energy. These are necessary for checking the scale-independence of the
effective potential. 
Section 
\ref{sec:example} treats a numerical example.

\section{Conventions and setup}
\label{sec:setup}
\setcounter{equation}{0}
\setcounter{footnote}{1}

In this section, I list the necessary facts and conventions used in this 
paper.
The MSSM is a softly-broken supersymmetric theory, with $SU(3)_c \times
SU(2)_L \times U(1)_Y$ gauge couplings $g_3$, $g$, and $\gp$. The last
is related to the GUT normalized coupling $g_1$ by
$\gp = \sqrt{{3/5}} \, g_1$. 
The superpotential of the MSSM is:
\beq
W = \overline u \Yu Q H_u - \overline d \Yd Q H_d -
\overline e \Ye L H_d + \mu H_u H_d,
\eeq
where $H_u$ and $H_d$ are the Higgs chiral superfields, and
the $Q,L$ are the chiral superfields containing left-handed quarks and
leptons and $u,d,e$ are those containing the conjugates of the 
right-handed quarks and leptons. The quark and lepton superfields carry
a suppressed family index, so that $\Yu$, $\Yd$, $\Ye$ are $3\times 3$
matrices in family space. Gauge indices for $SU(2)_L$ and $SU(3)_c$
are also suppressed, as in ref.~\cite{primer}. The Higgs
mass parameter $\mu$ can have an arbitrary phase.
The soft supersymmetry-breaking part of the Lagrangian is:
\beq
-{\cal L}_{\rm soft} &=& 
\Bigl ( {1\over 2} M_1 \stilde B \stilde B 
+ {1\over 2} M_2 \stilde W \stilde W 
+ {1\over 2} M_3 \stilde g \stilde g  
\nonumber \\
&&
+ \overline u \Au Q H_u - \overline d \Ad Q H_d -
\overline e \Ae L H_d + b H_u H_d \Bigr )
+ {\rm c.c.} 
\nonumber \\
&&
+ m^2_{H_u} |H_u|^2 + m^2_{H_d} |H_d|^2
\nonumber \\
&&
+ Q^\dagger \bm^2_Q Q + L^\dagger \bm^2_L L +
\overline u \bm^2_u \overline u^\dagger +
\overline d \bm^2_d \overline d^\dagger +
\overline e \bm^2_e \overline e^\dagger.
\eeq
where $\stilde B$, $\stilde W$, and $\stilde g$ are the 
bino, wino, and gluino fields, and the same symbols are used for scalar
fields as for the corresponding superfields. Here, $\Au$, $\Ad$, $\Ae$
are scalar cubic couplings in the form of $3\times 3$ 
matrices in family space. The Higgs fields have soft 
supersymmetry-breaking squared-mass
running parameters $m_{H_u}^2$, $m_{H_d}^2$, and $b$. The first two of
these are necessarily real, and by convention $b$ is taken to be real
at the renormalization scale $Q$ at which the effective potential is to be 
minimized.
This can always be achieved by a suitable field rephasing, and
ensures that the VEVs obtained by minimizing the full effective 
potential will also be real. The squarks and sleptons
have running soft supersymmetry-breaking squared masses 
$\bm^2_Q$, $\bm^2_L$, $\bm^2_u$, $\bm^2_d$, and $\bm^2_e$, which
are $3\times 3$ Hermitian matrices in flavor space.

The gauge-eigenstate complex scalar doublet
Higgs fields that are components of left-handed chiral superfields are 
called $H_u = (H_u^+,H_u^0)$ and $H_d = (H_d^0, H_d^-)$. The
electrically-neutral components have real positive VEVs 
$v_u$ and $v_d$ 
respectively. The field-dependent tree-level squared masses 
for vector bosons are then given by
\beq
m^2_{W} &=& {g^2 \over 2} (v_u^2 + v_d^2);
\\
m^2_{Z} &=& {g^2 + \gp^2 \over 2} (v_u^2 + v_d^2) .
\eeq
The tree-level Higgs potential is  
\beq
V\! &=&\! \Lambda +
(|\mu|^2 + m^2_{H_u}) (|H_u^0|^2 + |H_u^+|^2)
+ (|\mu|^2 + m^2_{H_d}) (|H_d^0|^2 + |H_d^-|^2)
\nonumber \\ &&+\, b\, (H_u^+ H_d^- - H_u^0 H_d^0) + \conj
\nonumber \\ &&+ {1 \over 8} (g^2 + g^{\prime 2})
( |H_u^0|^2 + |H_u^+|^2 - |H_d^0|^2 - |H_d^-|^2 )^2
+ {g^2 \over 2} |H_u^+ H_d^{0*} + H_u^0 H_d^{-*}|^2 .
\label{bighiggsv}
\eeq
Here $\Lambda$ is a running field-independent vacuum energy, which must be
included to maintain renormalization-scale independence of the effective
potential \cite{Einhorn:1983pp}-%
\cite{Ford:1993mv}.
The neutral Higgs scalar tree-level squared masses are obtained 
by
diagonalizing the matrices:
\beq
m^2_{\phi_R^0} &=&
\pmatrix{
|\mu|^2 + m^2_{H_u} + {g^2 + \gp^2 \over 4}(3 v_u^2 - v_d^2)
&
-b-({g^2 + \gp^2}) v_u v_d/2
\cr
-b-({g^2 + \gp^2}) v_u v_d/2
&
|\mu|^2 + m^2_{H_d} + {g^2 + \gp^2\over 4}(3 v_d^2 - v_u^2)
};
\\ 
m^2_{\phi_I^0} &=&
\pmatrix{
|\mu|^2 + m^2_{H_u} + {g^2 + \gp^2 \over 4}( v_u^2 - v_d^2)
& b 
\cr 
b &
|\mu|^2 + m^2_{H_d} + {g^2 + \gp^2 \over 4}( v_d^2 - v_u^2)
},
\eeq
\vskip5pt\noindent
which are written in the $({\rm Re}[H_u^0],{\rm Re}[H_d^0])$
and $({\rm Im}[H_u^0],{\rm Im}[H_d^0])$ bases, respectively.
The complex charge $\pm 1$ Higgs scalar tree-level squared masses are 
obtained
by diagonalizing the matrix
\beq
m^2_{\phi^\pm} =
\pmatrix{
|\mu|^2 + m^2_{H_u} 
+ {g^2 + \gp^2 \over 4}  v_u^2  
+ {g^2 - \gp^2 \over 4}  v_d^2 
& 
b + {g^2} v_u v_d/2
\cr
b + {g^2} v_u v_d/2 
&
|\mu|^2 + m^2_{H_d} 
+ {g^2 + \gp^2 \over 4}  v_d^2  
+ {g^2 - \gp^2 \over 4}  v_u^2 
},
\eeq
\vskip5pt
\noindent which is written in the $(H_u^+, H_d^{-*})$ basis.
The gauge-eigenstate fields can be expressed in terms of the tree-level 
squared-mass eigenstate fields as:
\beq
\pmatrix{H_u^0 \cr H_d^0}
&=& \pmatrix{v_u \cr v_d}
+ {1\over \sqrt{2}} { R}_\alpha \pmatrix{h^0 \cr H^0} 
+ {i\over \sqrt{2}} { R_{\beta_0}} \pmatrix{G^0 \cr A^0} 
\eeq
and
\beq
\pmatrix{H_u^+ \cr H_d^{-*}}
= {R}_{\beta_\pm} \pmatrix{G^+ \cr H^+},
\eeq
where $G^0$ and $G^\pm$ are Nambu-Goldstone fields,
$h^0, H^0, A^0$, and $H^\pm$ are the physical Higgs tree-level 
mass-eigenstate 
fields, and
\beq
{R}_\alpha = \pmatrix{ \calpha & \salpha \cr
                             -\salpha & \calpha};
\qquad\qquad\!\!\!
{R}_{\beta_0} = \pmatrix{ \sbetaO & \cbetaO \cr
                             -\cbetaO & \sbetaO};
\qquad\qquad\!\!\!
{R}_{\beta_\pm} = \pmatrix{ \sbetapm & \cbetapm \cr
                             -\cbetapm & \sbetapm}
\eeq
are orthogonal matrices determined by the requirements that:
\beq
{R}_\alpha^{-1} {m^2_{\phi_R^0}} R_\alpha 
&=& \pmatrix{m_{h^0}^2 & 0 \cr 0 & m_{H^0}^2};
\\
{R}_{\beta_0}^{-1} {m^2_{\phi_I^0}} R_{\beta_0} 
&=& \pmatrix{m_{G^0}^2 & 0 \cr 0 & m_{A^0}^2};
\\
R_{\beta_\pm}^{-1} {m^2_{\phi^\pm}} R_{\beta_\pm} 
&=& \pmatrix{m_{G^\pm}^2 & 0 \cr 0 & m_{H^\pm}^2} .
\eeq
These equations define the tree-level squared mass eigenvalues for
the Higgs scalar sector and the mixing angles $\alpha$, $\beta_0$ and
$\beta_\pm$.
Here I use a notation in which $c$ and $s$ with a subscript
indicate the cosine and sine of the indicated angle. So, in accord
with the orthogonality of the rotation matrices, $c_\alpha$ means
$\cos\alpha$, etc. Similarly, in the following, $c_{2\alpha}$ is taken to 
mean $\cos(2\alpha)$, and so on. Also, for future convenience, define 
coefficients
\beq
&k_{uh^0} = k_{d H^0} = \calpha;\qquad\qquad& 
k_{uH^0} = -k_{d h^0} = \salpha;
\\
&k_{uG^0} = k_{d A^0} = i\sbetaO;\qquad\qquad& 
k_{uA^0} = -k_{d G^0} = i\cbetaO;
\\
&k_{uG^+} = k_{d H^+} = \sbetapm;\qquad\qquad& 
k_{uH^+} = -k_{d G^+} = \cbetapm .
\eeq
Conventionally, $\calpha$, $\cbetaO$, and $\cbetapm$ are taken to be 
positive.
Because the minimum of the effective potential
is not a minimum of the tree-level potential, the 
angles $\beta_0$ and $\beta_\pm$ for the rotations in the pseudo-scalar
and charged Higgs sector are 
distinct from each other, and from $\beta = \tan^{-1}(v_u/v_d)$ at the 
minimum of the effective potential.
Some care should be used to distinguish these. Also, note that unlike the 
case in the ordinary Standard Model, $m_{G^0}^2 \not= m_{G^\pm}^2$ at 
tree level. Note also that even with arbitrary CP-violating phases,
there is no mixing in the tree-level squared masses of the neutral scalar 
$(h^0, H^0)$ and the pseudo-scalar $(G^0, A^0)$ sectors, because of the
freedom to choose the argument of $b$ to be 0 at any 
particular 
renormalization scale $Q$.

In section \ref{sec:potential}, $\displaystyle \sum_{\phi^0}$ and 
$\displaystyle \sum_{\phi^\pm}$ 
will appear, denoting sums over the lists
\beq
\phi^0 = (h^0,H^0,G^0,A^0)\qquad\mbox{and}\qquad \phi^\pm = (\Gpm,\Hpm)
\eeq 
respectively.

The neutralinos ($\Ni$; $i=1,2,3,4$) and charginos ($\stilde C_i^\pm$;
$i=1,2$) are mixtures of the electroweak gaugino and
Higgsino fields. In the $(\stilde B, \stilde W^0, \stilde H_d^0, \stilde
H_u^0)$ gauge-eigenstate basis, the neutralino mass matrix is
\beq
M_{\tilde N} = \pmatrix{M_1 & 0 & -\gp \vd/\sqrt{2} & \gp \vu/\sqrt{2} \cr
                         0 & M_2 & g \vd/\sqrt{2} & - g\vu /\sqrt{2} \cr
                        -\gp \vd/\sqrt{2} & g \vd/\sqrt{2} & 0 & -\mu \cr
                        \gp \vu/\sqrt{2} & - g\vu /\sqrt{2} & -\mu & 0}.
\eeq
This is diagonalized by a unitary matrix $N$:
\beq
N^* M_{\tilde N} N^{-1} &=& 
{\rm diag}( m_{\tilde N_1}, m_{\tilde N_2}, m_{\tilde N_3}, 
m_{\tilde N_4}),
\label{diagonalizemN}
\eeq
where the mass eigenvalues $m_{\tilde N_i}$ are all real and positive.
This can always be accomplished (for any phases of $M_1$, $M_2$, and 
$\mu$) by the following procedure. First, define $E$ to be a matrix
whose columns are orthonormal eigenvectors of the Hermitian 
squared-mass matrix
$M_{\tilde N}^\dagger M_{\tilde N}$, arranged in order of increasing
eigenvalue. The orthonormality means 
that $E^\dagger E=1$. It follows
that
\beq
E^T M_{\tilde N} E &=& P^2 m_D ,
\eeq
where $m_D$ is the matrix on the right-hand side of 
eq.~(\ref{diagonalizemN}),
and $P$ is a diagonal phase matrix. Then
\beq
N = P E^\dagger
\eeq
satisfies eq.~(\ref{diagonalizemN}), and also 
\beq
N M_{\tilde N}^\dagger M_{\tilde N} N^{-1} =
{\rm diag}( m^2_{\tilde N_1}, m^2_{\tilde N_2}, m^2_{\tilde N_3}, 
m^2_{\tilde N_4}) .
\eeq
This procedure gives the tree-level field-dependent neutralino squared 
masses and mixing matrix $N_{ij}$. (A suitable generalization of this 
method can be used to transform the mass and squared-mass matrices into a 
real positive diagonal form for any fermions in any theory.)

The tree-level chargino mass matrix is given by
\beq
M_{\tilde C} = \pmatrix{M_2 & g\vu \cr
                        g \vd & \mu},
\eeq
and is similarly diagonalized by unitary matrices $U$ and $V$ according 
to:
\beq
U^* M_{\tilde C} V^\dagger &=& 
\pmatrix{m_{\tilde C_1} & 0 \cr
           0 & m_{\tilde C_2}}
\eeq
where again $m_{\tilde C_i}$ are real and positive. Then
\beq
VM_{\tilde C}^\dagger M_{\tilde C}V^{-1} =
U M_{\tilde C}^* M_{\tilde C}^T U^{-1} = 
\pmatrix{m_{\tilde C_1}^2 & 0 \cr
           0 & m_{\tilde C_2}^2}.
\eeq

The tree-level squared mass of the gluino, denoted $m^2_{\tilde g} = 
|M_3|^2$,
does not depend on the VEVs $v_u$ and $v_d$.

In most of this paper, I will employ the approximation that only the 
third-family
Yukawa couplings $y_t$, $y_b$, and $y_\tau$ are significant, so that:
\beq
\Yu = \pmatrix{0&0&0\cr
               0&0&0\cr
               0&0&y_t};\qquad
\Yd = \pmatrix{0&0&0\cr
               0&0&0\cr
               0&0&y_b};\qquad
\Ye = \pmatrix{0&0&0\cr
               0&0&0\cr
               0&0&y_\tau}.
\eeq
By suitable
field redefinitions, they can be chosen real and positive.
The non-zero quark and lepton squared masses are therefore:
\beq
m_t^2 = y_t^2 \vu^2;\qquad\qquad
m_b^2 = y_b^2 \vd^2;\qquad\qquad
m_\tau^2 = y_\tau^2 \vd^2 .
\eeq
(Effects of the other Yukawa couplings are certainly smaller than the 
dominant 3-loop order contributions.) 

I will also assume here that the
soft supersymmetry-breaking scalar cubic 
couplings involving first- and second-family sfermions vanish,
so 
\beq
\Au = \pmatrix{0&0&0\cr
               0&0&0\cr
               0&0&a_t};\qquad
\Ad = \pmatrix{0&0&0\cr
               0&0&0\cr
               0&0&a_b};\qquad
\Ae = \pmatrix{0&0&0\cr
               0&0&0\cr
               0&0&a_\tau},
\eeq
and that, in the same basis,
the soft supersymmetry-breaking scalar
squared mass parameters are diagonal in family space, so
\beq
\bm^2_Q = \pmatrix{m^2_{Q_1} & 0 & 0 \cr
                    0 & m^2_{Q_2} & 0 \cr
                    0 & 0 & m^2_{Q_3} }, \>\>{\rm etc.}
\eeq
The parameters\footnote{In the literature, one often sees
rescaled quantities $A_t = a_t/y_t$, $A_b = a_b/y_t$, 
$A_\tau = a_\tau/y_\tau$, and $B = b/\mu$.} 
$a_t$, $a_b$, and $a_\tau$ 
can in general be complex,
and $m^2_{Q_{1,2,3}}$ etc.~are real soft running \DRbarprime squared 
mass parameters.
The squared
masses of the first-family squarks and sleptons are then given by:
\beq
m^2_{\tilde u_L}\!\! &=\> m^2_{Q_1} + \Delta_{\tilde u_L};
\qquad\qquad
m^2_{\tilde d_L}\!\! &=\> m^2_{Q_1} + \Delta_{\tilde d_L};
\label{msupdownL}
\\
m^2_{\tilde u_R}\!\! &=\> m^2_{u_1} + \Delta_{\tilde u_R};
\qquad\qquad
m^2_{\tilde d_R}\!\! &=\> m^2_{d_1} + \Delta_{\tilde d_R};
\\
m^2_{\tilde \nu_e}\!\! &=\> m^2_{L_1} + \Delta_{\tilde \nu_e};
\qquad\qquad
m^2_{\tilde e_L}\!\! &=\> m^2_{L_1} + \Delta_{\tilde e_L};
\\
m^2_{\tilde e_R}\!\! &=\> m^2_{e_1} + \Delta_{\tilde e_R},
\phantom{m^2_{\tilde \nu_e}}\qquad\qquad &
\phantom{=\> m^2_{L_1} + \Delta_{\tilde \nu_L};}
\label{mselR}
\eeq
The $D$-term contributions are:
\beq
\Delta_{\tilde f} &=& {1\over 2}
(I_{\tilde f}g^2 - Y_{\tilde f} \gp^2)(\vd^2 - \vu^2) ,
\eeq
with $I_{\tilde f}$ and $Y_{\tilde f}$ defined to be the
third component of weak isospin and the weak hypercharge of the
left-handed chiral superfield containing the squark or slepton $\stilde 
f$: 
\renewcommand{\arraystretch}{1.3}
\begin{center}
\begin{tabular}{l|rrrrrrr}
\phantom{xxx} 
& $\suL$ & $\sdL$ & $\snue$ & $\seL$ & $\suR$ & $\sdR$ & $\seR$
\\
\hline
$I_{\tilde f}$ & 
$\phantom{-}1/2$ & $-1/2$ & $1/2$ & $-1/2$ & $0$ & 
$\phantom{-1/}0$ & $\phantom{-1/}0$
\\
$Y_{\tilde f}$ & $1/6$ & $ 1/6$ & $-1/2$ & $-1/2$ & $-2/3$ & $1/3$ & $1$
\end{tabular}
\renewcommand{\arraystretch}{1.0}
\end{center}
Exactly analogous expressions hold for the second-family squark and
slepton squared masses 
$m^2_{\tilde c_L}$,
$m^2_{\tilde s_L}$,
$m^2_{\tilde c_R}$,
$m^2_{\tilde s_R}$,
$m^2_{\tilde \nu_\mu}$,
$m^2_{\tilde \mu_L}$, and
$m^2_{\tilde \mu_R}$, 
with the subscripts 1 replaced by 2
on the right-hand sides of eqs.~(\ref{msupdownL})-(\ref{mselR}).

For the stop, sbottom, stau, and stau-neutrino 
squared masses:
\beq
m^2_{\tilde t} &=&
\pmatrix{ m_{Q_3}^2 + y_t^2 \vu^2 + \Delta_{\tilde u_L} 
          & v_u a_t^* - v_d \mu y_t \cr
            v_u a_t - v_d \mu^* y_t & 
          m_{u_3}^2 + y_t^2 \vu^2 + \Delta_{\tilde u_R}};
\label{stops}
\\
m^2_{\tilde b} &=&
\pmatrix{ m_{Q_3}^2 + y_b^2 \vd^2 + \Delta_{\tilde d_L} 
        & v_d a_b^* - v_u \mu y_b \cr
          v_d a_b - v_u \mu^* y_b 
        & m_{d_3}^2 + y_b^2 \vd^2 + \Delta_{d_R}};
\label{sbots}
\\
m^2_{\tilde \tau} &=&
\pmatrix{ m_{L_3}^2 + y_\tau^2 \vd^2 + \Delta_{\tilde e_L} 
        & v_d a_\tau^* - v_u \mu y_\tau \cr
          v_d a_\tau - v_u \mu^* y_\tau & 
        m_{e_3}^2 + y_\tau^2 \vd^2 + \Delta_{\tilde e_R}};
\label{staus}
\\
m^2_{\tilde \nu_\tau} &=& m_{\tilde L_3}^2 + \Delta_{\tilde \nu_e}.
\eeq
Here the stop, sbottom and stau squared mass matrices are given 
in the $(\stilde f_L, \stilde f_R)$ bases. 
The eigenvalues of eqs.~(\ref{stops})-(\ref{staus}) give the squared
mass eigenvalues $m_{\tilde t_i}^2$, $m_{\tilde b_i}^2$, and
$m_{\tilde \tau_i}^2$ for $i=1,2$.
The squared-mass matrices are diagonalized by unitary transformations:
\beq
\pmatrix{\stilde f_L \cr \stilde f_R}
= X_{\tilde f}
\pmatrix{\stilde f_1 \cr \stilde f_2}
\eeq
for $f=t,b,\tau$, with
\beq
X_{\tilde f} = \pmatrix{ L_{\tilde f_1} & L_{\tilde f_2} \cr
            R_{\tilde f_1} & R_{\tilde f_2}},
\eeq
so that
\beq
X_{\tilde t}^{-1} m^2_{\tilde t} X_{\tilde t} =
\pmatrix{m_{\tilde t_1}^2 & 0 \cr
         0 & m^2_{\tilde t_2}},
\eeq
and similarly for the sbottoms and staus.  
Unitarity of the matrix $X_{\tilde f}$ allows one to write $L_{\tilde f_1} 
= R_{\tilde f_2}^* = 
c_{\tilde f}$, and 
$R_{\tilde f_1} = -L_{\tilde f_2}^* = s_{\tilde f}$, with
\beq
|c_{\tilde f}|^2 + |s_{\tilde f}|^2 = 1.
\eeq
(If the off-diagonal elements of the squared mass matrix are real, then
$c_{\tilde f}$ and $s_{\tilde f}$ are the sine and cosine of a
sfermion mixing angle.)
In the following, I will also make use of coefficients:
\beq 
&&x_{\tilde f} \,=\, I_{\tilde f},\qquad\qquad(\tilde f=\>
\mbox{first or second family sfermion});
\phantom{xxx}
\\ 
&&x_{\tilde t_i} = {1\over 2}|L_{\tilde t_i}|^2;\qquad\>\>\>\>\>
x_{\tilde b_i} = -{1\over 2}|L_{\tilde b_i}|^2;\qquad\>\>\>\>\>
x_{\tilde \nu_\tau} = {1\over 2}; \qquad\>\>\>\>\>
x_{\tilde \tau_i} = -{1\over 2}|L_{\tilde \tau_i}|^2,
\phantom{xxx}
\eeq
and
\beq
&& x'_{\tilde f} \,=\, Y_{\tilde f},\qquad\qquad(\tilde f=\>
\mbox{first or second family sfermion});
\\
&& x'_{\tilde t_i} = 
{1\over 6} |L_{\tilde t_i}|^2 - {2\over 3} |R_{\tilde t_i}|^2;
\qquad\qquad\>\>\>
x'_{\tilde b_i} = 
{1\over 6} |L_{\tilde b_i}|^2 + {1\over 3} |R_{\tilde b_i}|^2;
\\
&& x'_{\tilde \tau_i} = 
-{1\over 2} |L_{\tilde \tau_i}|^2 + |R_{\tilde \tau_i}|^2;
\qquad\qquad\>\>\>
x'_{\tilde \nu_\tau} = -{1\over 2}.
\eeq
When $\displaystyle \sum_{\tilde f}$ appears in section 
\ref{sec:potential}, it will denote a sum 
over the list
\beq
\tilde f = (
\suL, \sdL, \suR, \sdR,
\snue, \seL, \seR, 
\scL, \ssL, \scR, \ssR,
\snumu, \smuL, \smuR, 
\stopI, \stopII, \sbotI, \sbotII,
\snutau, \stauI, \stauII 
).
\eeq
The symbol 
\beq
n_{\tilde f} = 
\left \{ \begin{array}{ll}
1, & \tilde f =\>\mbox{slepton} \\
3, & \tilde f =\>\mbox{squark}   
\end{array} \right.
\eeq
denotes the number of colors.

To summarize, evaluation of the effective potential can proceed as 
follows.
At a renormalization scale $Q$, choose values for the set of input data 
consisting of the 33 \DRbarprime 
running 
parameters:
\bea
&&v_u,\> v_d,\\
&&g_3, \>g,\> \gp,\> y_t,\> y_b,\> y_\tau, \\
&&m_{Q_i}^2,\> m_{L_i}^2,\> m_{u_i}^2,\> m_{d_i}^2,\> m_{e_i}^2,\qquad 
(i=1,2,3)\\
&& m_{H_u}^2,\> m_{H_d}^2,\> b,\> \mu, \\
&& M_3,\> M_2,\> M_1,\> a_t,\> a_b,\> a_\tau, 
\eea
of which the last 7 may be complex.
Using the preceding protocols, evaluate the mixing parameters
\beq
&&\alpha, \>\beta_0,\> \beta_\pm,\\
&& N_{ij},\> U_{ij},\> V_{ij},\\
&&L_{\tilde t_i},
\>
R_{\tilde t_i},
\>
L_{\tilde b_i},
\>
R_{\tilde b_i},
\>
L_{\tilde \tau_i},
\>
R_{\tilde \tau_i},\phantom{xxxxxxx}
\eeq
and the 
39 distinct field-dependent 
tree-level squared masses:
\bea
&&m_W^2, 
\>
m_Z^2,
\\
&&m_{h^0}^2, 
\>
m_{H^0}^2, 
\>
m_{G^0}^2, 
\>
m_{A^0}^2, 
\>
m_{G^\pm}^2, 
\>
m_{H^\pm}^2, 
\\ 
&&m_{\tilde g}^2, 
\>
m^2_{\tilde N_i},
\>
m^2_{\tilde C_i},
\\
&&m_t^2, m_b^2, m_\tau^2,
\\
&&m_{\tilde u_L}^2, 
\>
m_{\tilde d_L}^2, 
\>
m_{\tilde u_R}^2, 
\>
m_{\tilde d_R}^2, 
\>
m_{\tilde \nu_e}^2, 
\>
m_{\tilde e_L}^2, 
\>
m_{\tilde e_R}^2, 
\\
&&m_{\tilde c_L}^2, 
\>
m_{\tilde s_L}^2, 
\>
m_{\tilde c_R}^2, 
\>
m_{\tilde s_R}^2, 
\>
m_{\tilde \nu_\mu}^2, 
\>
m_{\tilde \mu_L}^2, 
\>
m_{\tilde \mu_R}^2, 
\\
&&m_{\tilde t_1}^2,
\>
m_{\tilde t_2}^2,
\>
m_{\tilde b_1}^2,
\>
m_{\tilde b_2}^2,
\>
m_{\tilde \nu_\tau}^2 ,
\>
m_{\tilde \tau_1}^2,
\>
m_{\tilde \tau_2}^2.
\eea
These squared masses (and implicitly $Q$) then become arguments for the 
functions
$\FSSS$, $\FSS$, $\FFFS$, $\FFFbS$, $\FSSV$, 
$\FVS$, $\FVVS$, $\FFFV$, $\FFFbV$, and $\Fgauge$
appearing in the next section, and defined explicitly in 
eqs.~(2.17)-(2.22), (4.12)-(4.16), and (6.21)-(6.30) of
ref.~\cite{general}. (For the sake of uniformity of notation, I use
the symbols $\FSSS$, $\FSS$, $\FFFS$, $\FFFbS$, and $\FSSV$ in place of 
$f_{SSS}$, $f_{SS}$, $f_{FFS}$, $f_{\overline{FF}S}$, and $f_{SSV}$
for the functions which are the same in the \MSbar and \DRbarprime
schemes. All two-loop functions used in the present paper are 
\DRbarprime ones.)
To simplify the notation, I adopt the
convention that the name of a particle is synonymous with its
squared mass when appearing as an argument of one of these functions.
So, for example, $\FSSS(h^0,A^0,G^0)$ means $\FSSS(m_{h^0}^2, m_{A^0}^2,
m_{G_0}^2)$.

\section{MSSM effective potential}
\label{sec:potential}
\setcounter{equation}{0}
\setcounter{footnote}{1}

\subsection{Tree-level and one-loop contributions}
\label{subsec:oneloop}

The tree-level contribution to the effective potential for $v_u$ and $v_d$ 
is:
\beq
V^{(0)} = \Lambda + 
(|\mu|^2 + m^2_{H_u}) v_u^2 + (|\mu|^2 + m^2_{H_d}) v_d^2 
- 2 b v_u v_d 
+{1\over 8}(g^2 + \gp^2) (v_u^2 - v_d^2)^2.
\eeq
The one-loop contribution in the \DRbarprime scheme is:
\beq
V^{(1)} &=& \sum_{\phi^0} h(\phi^0) +
2 \sum_{\phi^\pm} h(\phi^\pm) +
2 \sum_{\tilde f} n_{\tilde f} h(\tilde f)
-2 \sum_{i=1}^4 h(\Ni)
-4 \sum_{i=1}^2 h(\Ci)
\nonumber \\ &&
- 16 h(\stilde g)
-12  h(t) - 12 h(b) - 4 h(\tau)
+ 3 h(Z) + 6 h(W),
\eeq
where $h(x)$ is the function defined in eq.~(\ref{defineh}),
and the name of each particle is used to denote its squared mass.

\subsection{$SSS$-diagram contributions}
\label{subsec:SSS}

In this subsection, I list the contributions to the MSSM two-loop 
effective potential from diagrams with three scalar propagators.
These all involve the function $\FSSS(x,y,z)$, with arguments $x,y,z$
equal to tree-level field-dependent scalar squared masses.

The contributions from diagrams with three Higgs scalar propagators are:
\beq
V^{(2)}_{\phi^0\phi^0\phi^0} &=& 
{(g^2 + \gp^2)^2 \over 32} \Biggl [
(\calpha v_u + \salpha v_d)^2 
\biggl \lbrace
3 \Calpha^2 \FSSS(h^0,h^0,h^0) 
+ 2 \SbetaO^2 \FSSS (h^0,A^0,G^0) 
\nonumber \\ &&
+\CbetaO^2 [ \FSSS(h^0,G^0,G^0) + \FSSS(h^0,A^0,A^0) ]
\biggr\rbrace
\nonumber \\ &&
+ (\salpha v_u - \calpha v_d)^2 
\biggl \lbrace
3 \Calpha^2 \FSSS(H^0,H^0,H^0) 
+ 2 \SbetaO^2 \FSSS (H^0,A^0,G^0) 
\nonumber \\ &&
+\CbetaO^2 [ \FSSS(H^0,G^0,G^0) + \FSSS(H^0,A^0,A^0) ]
\biggr\rbrace
\nonumber \\ &&
+ [\salpha (1 - 6 \calpha^2) v_u + \calpha (1-6 \salpha^2 ) v_d]^2
\FSSS(h^0,h^0,H^0)
\nonumber \\ &&
+ [\salpha (1 - 6 \calpha^2) v_d - \calpha (1-6 \salpha^2 ) v_u]^2
\FSSS(h^0,H^0,H^0)
\Biggr ]
\eeq
and 
\beq
&& V^{(2)}_{\phi^0 \phi^\pm \phi^\pm} = {1\over 16} \Biggl \lbrace
2\Bigl [g^2 \Cbetapm (\salpha v_u - \calpha v_d) +
\gp^2 \Sbetapm (\calpha v_u + \salpha v_d) \Bigr ]^2
\FSSS (h^0, G^\pm , H^\pm) 
\nonumber \\ && 
+ 2 \Bigl [g^2 \Cbetapm (\calpha v_u + \salpha v_d) -
\gp^2 \Sbetapm (\salpha v_u - \calpha v_d) \Bigr ]^2
\FSSS (H^0, G^\pm , H^\pm) 
\nonumber \\ && 
+ \Bigl [
g^2 (\calpha v_u - \salpha v_d) 
+ g^2 \Sbetapm (\salpha v_u - \calpha v_d) 
- \gp^2 \Cbetapm (\calpha v_u + \salpha v_d) \Bigr ]^2 
\FSSS(h^0, G^\pm, G^\pm)
\nonumber \\ && 
+ \Bigl [
g^2 (\calpha v_u - \salpha v_d) 
-g^2 \Sbetapm (\salpha v_u - \calpha v_d) 
+ \gp^2 \Cbetapm (\calpha v_u + \salpha v_d) \Bigr ]^2 
\FSSS(h^0, H^\pm, H^\pm) 
\nonumber \\ && 
+ \Bigl [
g^2 (\salpha v_u + \calpha v_d) 
- g^2 \Sbetapm (\calpha v_u + \salpha v_d) 
- \gp^2 \Cbetapm (\salpha v_u - \calpha v_d) \Bigr ]^2 
\FSSS(H^0, G^\pm, G^\pm)
\nonumber \\ && 
+ \Bigl [
g^2 (\salpha v_u + \calpha v_d) 
+ g^2 \Sbetapm (\calpha v_u + \salpha v_d) 
+ \gp^2 \Cbetapm (\salpha v_u - \calpha v_d) \Bigr ]^2 
\FSSS(H^0, H^\pm, H^\pm)
\nonumber \\ && 
+ 2 g^4 (\cbetaO v_u - \sbetaO v_d)^2 \FSSS(G^0,G^\pm,H^\pm)
+ 2 g^4 (\sbetaO v_u + \cbetaO v_d)^2 \FSSS(A^0,G^\pm,H^\pm)
\Biggr \rbrace .
\eeq

The contributions from diagrams with two sfermions and a neutral
Higgs scalar are:
\beq
V^{(2)}_{ \phi^0\tilde f \tilde f'} &=& 
{1\over 2} \sum_{\phi^0, \tilde f, \tilde f^{\prime}}
n_{\tilde f} |\lambda_{\phi^0 \tilde f \tilde f^{\prime *}}|^2
\FSSS(\phi^0, \stilde f, \stilde f') .
\label{Vsfsfphizero}
\eeq
The couplings involving first- and second-family sfermions, and
the tau sneutrino, are 
non-zero
only when $\stilde f = \stilde f'$. They are given by:
\beq
\lambda_{\phi^0 \tilde f \tilde f^*} = {1 \over \sqrt{2}}
\left ( I_{\tilde f} g^2  - Y_{\tilde f} \gp^2 \right )
{\rm Re}[ k_{u\phi^0} v_u - k_{d \phi^0} v_d],
\eeq
with $I_{\tilde f}, Y_{\tilde f}$ defined in section
\ref{sec:setup}.
The remaining couplings for the third-family sfermions are:
\beq
\lambda_{\phi^0 \tilde t_i \tilde t_j^*} &=&
L_{\tilde t_i} L^*_{\tilde t_j}
\lambda_{\phi^0 \tilde u_L \tilde u_L^*}
+ R_{\tilde t_i} R^*_{\tilde t_j}
\lambda_{\phi^0 \tilde u_R \tilde u_R^*}
- \sqrt{2} v_u y_t^2 {\rm Re}[k_{u\phi^0}] (L_{\tilde t_i} L^*_{\tilde t_j}
+ R_{\tilde t_i} R^*_{\tilde t_j} )
\nonumber \\ &&
- {1 \over \sqrt{2}} (k_{u\phi^0} a_t - k_{d\phi^0}^* \mu^* y_t)
L_{\tilde t_i} R^*_{\tilde t_j}
- {1 \over \sqrt{2}} (k_{u\phi^0}^* a_t^* - k_{d\phi^0} \mu y_t)
R_{\tilde t_i} L^*_{\tilde t_j};
\\
\lambda_{\phi^0 \tilde b_i \tilde b_j^*} &=&
L_{\tilde b_i} L^*_{\tilde b_j}
\lambda_{\phi^0 \tilde d_L \tilde d_L^*}
+ R_{\tilde b_i} R^*_{\tilde b_j}
\lambda_{\phi^0 \tilde d_R \tilde d_R^*}
- \sqrt{2} v_d y_b^2 {\rm Re}[k_{d\phi^0}] (
L_{\tilde b_i} L^*_{\tilde b_j}
+ R_{\tilde b_i} R^*_{\tilde b_j} )
\nonumber \\ &&
- {1 \over \sqrt{2}} (k_{d\phi^0} a_b - k_{u\phi^0}^* \mu^* y_b)
L_{\tilde b_i} R^*_{\tilde b_j}
- {1 \over \sqrt{2}} (k_{d\phi^0}^* a_b^* - k_{u\phi^0} \mu y_b)
R_{\tilde b_i} L^*_{\tilde b_j};
\\
\lambda_{\phi^0 \tilde \tau_i \tilde \tau_j^*} &=&
L_{\tilde \tau_i} L^*_{\tilde \tau_j}
\lambda_{\phi^0 \tilde e_L \tilde e_L^*}
+ R_{\tilde \tau_i} R^*_{\tilde \tau_j}
\lambda_{\phi^0 \tilde e_R \tilde e_R^*}
- \sqrt{2} v_d y_\tau^2 {\rm Re}[k_{d\phi^0}] (
L_{\tilde \tau_i} L_{\tilde \tau_j}^*
+ R_{\tilde \tau_i} R^*_{\tilde \tau_j} )
\nonumber \\ &&
- {1 \over \sqrt{2}} (k_{d\phi^0} a_\tau - k_{u\phi^0}^* \mu^* y_\tau)
L_{\tilde \tau_i} R^*_{\tilde \tau_j}
- {1 \over \sqrt{2}} (k_{d\phi^0}^* a_\tau^* - k_{u\phi^0} \mu y_\tau)
R_{\tilde \tau_i} L^*_{\tilde \tau_j} .
\eeq

The contributions from diagrams with two sfermions and a charged Higgs
scalar are:
\beq
V^{(2)}_{\phi^\pm\tilde f \tilde f'} &=&
\sum_{\phi^+,\tilde f, \tilde f^{\prime}}
n_{\tilde f} |\lambda_{\phi^+ \tilde f \tilde f^{\prime *}}|^2
\FSSS(\phi^\pm, \stilde f, \stilde f').
\label{Vsfsfphipm}
\eeq
Here the non-zero couplings involving the first- and second-family 
sfermions are:
\beq
\lambda_{\phi^+ \tilde e_L \tilde \nu_e^*} = 
\lambda_{\phi^+ \tilde \mu_L \tilde \nu_\mu^*} =
\lambda_{\phi^+ \tilde d_L \tilde u_L^*} =
\lambda_{\phi^+ \tilde s_L \tilde c_L^*} &=&
-{g^2 \over 2} (k_{u\phi^+} v_u + k_{d\phi^+} v_d) ,
\eeq
and those involving the third-family sfermions are:
\beq
\lambda_{\phi^+ \tilde b_i \tilde t_j^*} &=&
L_{\tilde b_i} L^*_{\tilde t_j} \left (
\lambda_{\phi^+ \tilde d_L \tilde u_L^*}
+y_t^2 v_u k_{u\phi^+}
+y_b^2 v_d k_{d\phi^+}
\right )
+ R_{\tilde b_i} R^*_{\tilde t_j} 
y_t y_b (k_{d\phi^+} v_u + k_{u \phi^+}v_d )
\nonumber \\
&&
+ L_{\tilde b_i} R^*_{\tilde t_j} (k_{u\phi^+} a_t + k_{d\phi^+} \mu^* y_t)
+ R_{\tilde b_i} L^*_{\tilde t_j} (k_{d\phi^+} a_b^* + k_{u\phi^+} \mu y_b);
\\
\lambda_{\phi^+ \tilde \tau_i \tilde \nu_\tau^*} &=&
L_{\tilde \tau_i} \left (
\lambda_{\phi^+ \tilde e_L \tilde \nu_e^*}
+y_\tau^2 v_d k_{d\phi^+}
\right )
+ R_{\tilde \tau_i} (k_{d\phi^+} a_\tau^* +
k_{u\phi^+} \mu y_\tau) .
\eeq
 
\subsection{$SS$-diagram contributions}
\label{subsec:SS}

In this subsection, I list the contributions to the MSSM two-loop 
effective 
potential which come from diagrams with two scalar field propagators.
They all involve the function $\FSS(x,y)$.

The contributions proportional to $g_3^2$ are:
\beq
V^{(2), g_3^2}_{\tilde q \tilde q} &=& 2 g_3^2 \biggl \lbrace
\sum_{i,j=1}^2 \left [
|L_{\tilde t_i} L_{\tilde t_j}^* - R_{\tilde t_i} R_{\tilde t_j}^*|^2
\FSS(\stopi,\stopj)
+
|L_{\tilde b_i} L_{\tilde b_j}^* - R_{\tilde b_i} R_{\tilde b_j}^*|^2
\FSS(\sboti,\sbotj) \right ]
\nonumber \\ &&
+ \FSS (\suL,\suL) + \FSS (\suR,\suR) + \FSS(\sdL,\sdL) +\FSS (\sdR,\sdR)
\nonumber \\ &&
+\FSS (\scL,\scL) + \FSS (\scR,\scR) + \FSS(\ssL,\ssL) +\FSS (\ssR,\ssR)
\biggr \rbrace .
\eeq

The contributions from diagrams with two Higgs scalar propagators are:
\beq
V^{(2)}_{\phi^0\phi^0} &=& {g^2 + \gp^2 \over 32} \biggl \lbrace
3 \Calpha^2 [\FSS (h^0,h^0) + \FSS (H^0,H^0) ]
+ 3 \CbetaO^2 [\FSS (A^0,A^0) + \FSS (G^0,G^0)]
\nonumber \\ &&
+(4-6 \Calpha^2  ) \FSS (h^0, H^0)
+(4-6 \CbetaO^2  ) \FSS (A^0, G^0)
\nonumber \\ &&
+ 2 \Calpha \CbetaO
[ \FSS (h^0,A^0) + \FSS (H^0, G^0) - \FSS (h^0, G^0) - \FSS (H^0,A^0)]
\biggr \rbrace;
\\
V^{(2)}_{\phi^\pm\phi^\pm} &=& {g^2 + \gp^2 \over 4} \left \lbrace
\Cbetapm^2 [\FSS (\Hpm,\Hpm) + \FSS (\Gpm,\Gpm) ]
+(1- 2 \Cbetapm^2) \FSS (\Hpm, \Gpm ) \right \rbrace;\phantom{xxxx}
\\
V^{(2)}_{\phi^0\phi^\pm} &=&
{1\over 8} \biggl \lbrace
[g^2 (1 - \Salpha \Sbetapm) + \gp^2 \Calpha \Cbetapm ]
[\FSS(h^0,H^\pm) + \FSS(H^0, G^\pm)]
\nonumber \\ &&
+[g^2 (1 + \Salpha \Sbetapm) - \gp^2 \Calpha \Cbetapm ]
[\FSS(h^0,G^\pm) + \FSS(H^0, H^\pm)]
\nonumber \\ &&
+[g^2 (1 - \SbetaO \Sbetapm) + \gp^2 \CbetaO \Cbetapm ]
[\FSS(A^0,H^\pm) + \FSS(G^0, G^\pm)]
\nonumber \\ &&
+[g^2 (1 + \SbetaO \Sbetapm) - \gp^2 \CbetaO \Cbetapm ]
[\FSS(A^0,G^\pm) + \FSS(G^0, H^\pm)]
\biggr\rbrace .
\eeq

The contributions from diagrams involving electroweak gauge couplings
and one 
sfermion and one Higgs
scalar propagator are:
\beq
V^{(2),g^2,\gp^2}_{\tilde f \phi^0} &=& 
{1 \over 4} \sum_{\tilde f}\sum_{\phi^0} n_{\tilde f} 
(x_{\tilde f} g^2 - x'_{\tilde f} \gp^2)
(|k_{d\phi^0}|^2 - |k_{u\phi^0}|^2) \FSS (\stilde f, \phi^0);
\\
V^{(2),g^2,\gp^2}_{\tilde f \phi^\pm} &=& 
{1 \over 2} \sum_{\tilde f}\sum_{\phi^\pm} n_{\tilde f} 
(x_{\tilde f} g^2 + x'_{\tilde f} \gp^2)
(k_{u\phi^+}^2 - k_{d\phi^+}^2) \FSS (\stilde f, \phi^\pm).
\eeq

The contributions from diagrams involving electroweak gauge couplings
and two sfermion propagators are:
\beq
V^{(2),g^2}_{\tilde f \tilde f} &=& {g^2 \over 2}\biggl \lbrace
\sum_{\tilde f,\tilde f'}
n_{\tilde f} n_{\tilde f'} x_{\tilde f} x_{\tilde f'} \FSS(\stilde f,\stilde f')
+ \sum_{\tilde f} n_{\tilde f} x_{\tilde f}^2 \FSS(\stilde f,\stilde f)
\nonumber \\ &&
+ {3 \over 2} |L_{\tilde t_1} L_{\tilde t_2}|^2 \FSS(\stopI,\stopII) 
+ {3 \over 2} |L_{\tilde b_1} L_{\tilde b_2}|^2 \FSS (\sbotI,\sbotII)
+ {1 \over 2} |L_{\tilde \tau_1} L_{\tilde \tau_2}|^2 \FSS (\stauI,\stauII)
\nonumber \\ &&
+ 3 \Bigl [\FSS(\tilde u_L,\tilde d_L) + \FSS (\tilde c_L,\tilde s_L) + \sum_{i,j=1}^2
|L_{\tilde t_i} L_{\tilde b_j}|^2 \FSS (\stopi,\sbotj) \Bigr ]
\nonumber \\ &&
+ \FSS(\tilde e_L,\tilde \nu_e) + \FSS (\tilde \mu_L,\tilde \nu_\mu) +
\sum_{i=1}^2 |L_{\tilde \tau_i}|^2 \FSS (\staui,\tilde \nu_\tau)
\biggr \rbrace;
\\
V^{(2), \gp^2}_{\tilde f \tilde f} &=& {\gp^2 \over 2}  \biggl [
\sum_{\tilde f , \tilde f'} 
n_{\tilde f} n_{\tilde f'} x'_{\tilde f} x'_{\tilde f'}
\FSS(\tilde f, \tilde f') 
+ \sum_{\tilde f} 
n_{\tilde f} x_{\tilde f}^{\prime 2} \FSS(\tilde f,\tilde f)
\nonumber \\ && 
+ {1 \over 6}|L_{\tilde t_1} L_{\tilde t_2}^* - 4 R_{\tilde t_1} R_{\tilde t_2}^*|^2 
\FSS(\stopI,\stopII)
+ {1\over 6} |L_{\tilde b_1} L_{\tilde b_2}^* + 2 R_{\tilde b_1} R_{\tilde b_2}^*|^2 
\FSS(\sbotI,\sbotII)
\nonumber \\ && 
+ {1\over 2}| L_{\tilde \tau_1} L_{\tilde \tau_2}^* - 2 R_{\tilde \tau_1} R_{\tilde \tau_2}^*|^2 
\FSS(\stauI,\stauII)
\biggr ] .
\eeq
The contributions from diagrams involving Yukawa couplings and
one sfermion and one Higgs scalar propagator are:
\beq
V^{(2),y^2}_{\tilde f\phi^0} &=&
\sum_{\phi^0} \sum_{i=1}^2 \biggl \lbrace 
{3\over 2} y_t^2 |k_{u\phi^0}|^2 
\FSS(\stopi, \phi^0)
+{3\over 2} y_b^2 |k_{d\phi^0}|^2 
\FSS(\sboti, \phi^0)
+ {y_\tau^2\over 2} |k_{d\phi^0}|^2 
\FSS(\staui, \phi^0)
\biggr \rbrace;\phantom{xxxx}
\label{Vsfphiyy}
\\
V^{(2),y^2}_{\tilde f\phi^\pm} &\!=\!&
\sum_{\phi^\pm}\biggl [
\sum_{i=1}^2 \Bigl \lbrace
3(y_t^2 |R_{\tilde t_i}|^2 k_{u\phi^+}^2 + y_b^2 |L_{\tilde t_i}|^2 
k_{d\phi^+}^2 )
\FSS (\stopi, \phi^\pm)
+ y_\tau^2 |R_{\tilde \tau_i}|^2 k_{d\phi^\pm}^2 \FSS(\staui,\phi^\pm) 
\nonumber \\ &&
+ 3(y_t^2 |L_{\tilde b_i}|^2 k_{u\phi^+}^2 + y_b^2 |R_{\tilde b_i}|^2 
k_{d\phi^+}^2 )
\FSS (\sboti, \phi^\pm)
\Bigr \rbrace
+ y_\tau^2 k_{d\phi^\pm}^2 \FSS (\snutau, \phi^\pm)
\biggr ].\phantom{xx}
\eeq
The contributions from diagrams involving Yukawa couplings and
two sfermion propagators are:
\beq
V^{(2),y^2}_{\tilde f\tilde f} &=&
\sum_{i,j=1}^2 \biggl \lbrace
y_t^2 \bigl (  3 |L_{\tilde t_i} R_{\tilde t_j}|^2 
+ 9 L_{\tilde t_i} R_{\tilde t_i}^* R_{\tilde t_j} L_{\tilde t_j}^* \bigr ) \FSS(\stopi,\stopj)
\nonumber \\ &&
+ y_b^2 \bigl ( 3 |L_{\tilde b_i} R_{\tilde b_j}|^2 
+ 9 L_{\tilde b_i} R_{\tilde b_i}^* R_{\tilde b_j} L_{\tilde b_j}^* \bigr ) \FSS(\sboti,\sbotj)
+ {y_\tau^2 \over 2} |L_{\tilde \tau_i} R_{\tilde \tau_j} + 
           R_{\tilde \tau_i} L_{\tilde \tau_j}|^2 
\FSS(\staui,\stauj)
\nonumber \\ &&
+ 3 (y_t^2 |R_{\tilde t_i} L_{\tilde b_j}|^2 + y_b^2 |L_{\tilde t_i} R_{\tilde b_j}|^2)
\FSS (\stopi,\sbotj) 
+ 6 y_b y_\tau {\rm Re}[L_{\tilde b_i} R^*_{\tilde b_i} R_{\tilde \tau_j} L^*_{\tilde \tau_j}]
\FSS(\sboti, \stauj)
\biggr \rbrace
\nonumber \\ &&
+ y_\tau^2 \sum_{i=1}^2 |R_{\tilde \tau_i}|^2 \FSS(\staui,\stilde 
\nu_\tau) .
\label{Vsfsfyy}
\eeq

\subsection{$FFS$-diagram and  $\FFbS$-diagram contributions}
\label{subsec:FFS}

In this subsection I list the contributions to the MSSM two-loop effective
potential from diagrams involving scalars and fermions. These
include the $FFS$ diagrams (without chirality-flipping fermion mass 
insertions) and the
$\FFbS$ diagrams (which do have two such fermion mass insertions on 
different propagators).
They therefore involve the functions $\FFFS(x,y,z)$ and $\FFFbS(x,y,z)$.

The contributions from diagrams involving the gluino are given by:
\beq
V^{(2)}_{q \tilde g \tilde q}
&=& 8 g_3^2 \biggl \lbrace
\sum_{i=1}^2 \Bigl [
\FFFS (t, \stilde g, \stopi) - 2 {\rm Re}[L_{\tilde t_i} R_{\tilde t_i}^*] 
m_t m_{\tilde g}
\FFFbS (t, \stilde g, \stopi)
\nonumber \\ && 
+ \FFFS (b, \stilde g, \sboti) - 2 {\rm Re}[L_{\tilde b_i} R_{\tilde b_i}^*] 
m_b m_{\tilde g}
\FFFbS (b, \stilde g, \sboti) \Bigr ]
\nonumber \\ && 
+\FFFS (0,\stilde g,\suL)
+  \FFFS (0,\stilde g,\sdL)
+  \FFFS (0,\stilde g,\suR)
+  \FFFS (0,\stilde g,\sdR)
\nonumber \\ && 
+  \FFFS (0,\stilde g,\scL)
+  \FFFS (0,\stilde g,\ssL)
+  \FFFS (0,\stilde g,\scR)
+  \FFFS (0,\stilde g,\ssR)
\biggr\rbrace .
\phantom{xxx}
\eeq

The contributions from diagrams involving neutralinos, chiral fermions,
and sfermions 
are:
\beq
V^{(2)}_{f \tilde N \tilde f}
&=& \sum_{i=1}^4 \Biggl \lbrace \sum_{j=1}^2 \biggl [
3 (|Y_{t\tilde N_i \tilde t^*_j}|^2 +
|Y_{\overline t\tilde N_i \tilde t_j}|^2) \FFFS(t,\Ni, \stopj)
+ 6 {\rm Re}[
Y_{t\tilde N_i \tilde t^*_j} Y_{\overline t\tilde N_i \tilde t_j}]
m_t m_{\Ni} \FFFbS(t,\Ni, \stopj)
\nonumber \\ &&
+ 3 (|Y_{b\tilde N_i \tilde b^*_j}|^2 +
|Y_{\overline b\tilde N_i \tilde b_j}|^2) \FFFS(b,\Ni, \sbotj)
+ 6 {\rm Re}[
Y_{b\tilde N_i \tilde b^*_j} Y_{\overline b\tilde N_i \tilde b_j}]
m_b m_{\Ni} \FFFbS(b,\Ni, \sbotj)
\nonumber \\ &&
+  (|Y_{\tau\tilde N_i \tilde \tau^*_j}|^2 +
|Y_{\overline \tau\tilde N_i \tilde \tau_j}|^2) \FFFS(\tau,\Ni, \stauj)
+ 2 {\rm Re}[ Y_{\tau\tilde N_i \tilde \tau^*_j}
Y_{\overline \tau\tilde N_i \tilde\tau_j}]
m_\tau m_{\Ni} \FFFbS(\tau,\Ni, \stauj) \biggr ]
\nonumber \\ &&
+ |Y_{\nu \tilde N_i \tilde \nu^*} |^2
[\FFFS(0,\Ni, \snue) + \FFFS(0,\Ni, \snumu)+\FFFS(0,\Ni, \snutau)]
\nonumber \\ &&
+ 3 |Y_{u \tilde N_i \tilde u_L^*} |^2 
[\FFFS(0,\Ni, \suL) + \FFFS(0,\Ni, \scL)]
\nonumber \\ &&
+ 3 |Y_{\overline u \tilde N_i \tilde u_R} |^2 
[\FFFS(0,\Ni, \suR) + \FFFS(0,\Ni, \scR)]
\nonumber \\ &&
+ 3 |Y_{d \tilde N_i \tilde d_L^*} |^2 
[\FFFS(0,\Ni, \sdL) + \FFFS(0,\Ni, \ssL)]
\nonumber \\ &&
+ 3 |Y_{\overline d \tilde N_i \tilde d_R} |^2 
[\FFFS(0,\Ni, \sdR) + \FFFS(0,\Ni, \ssR)]
\nonumber \\ &&
+ |Y_{e \tilde N_i \tilde e_L^*} |^2 
[\FFFS(0,\Ni, \seL) + \FFFS(0,\Ni, \smuL)]
\nonumber \\ &&
+ |Y_{\overline e \tilde N_i \tilde e_R} |^2 
[\FFFS(0,\Ni, \seR) + \FFFS(0,\Ni, \smuR)]
\Biggr \rbrace .
\label{VfsfN}
\eeq
Here, the fermion-neutralino-sfermion couplings are: 
\beq
&&Y_{u\tilde N_i \tilde u^*_L}
= -{1\over \sqrt{2}} (g N_{i2}^* + {\gp\over 3} N_{i1}^*);
\qquad\qquad
Y_{\overline u\tilde N_i \tilde u_R}
= {2 \sqrt{2} \over 3} \gp N_{i1}^*;
\\
&&Y_{d\tilde N_i \tilde d^*_L}
= {1\over \sqrt{2}} (g N_{i2}^* - {\gp\over 3} N_{i1}^*);
\qquad\qquad\>\>\>\,
Y_{\overline d\tilde N_i \tilde d_R}
= -{\sqrt{2} \over 3} \gp N_{i1}^*;
\\
&&Y_{e\tilde N_i \tilde e^*_L}
= {1\over \sqrt{2}} (g N_{i2}^* + \gp N_{i1}^*);
\qquad\qquad\>\>\>\,
Y_{\overline e\tilde N_i \tilde e_R}
= -\sqrt{2} \gp N_{i1}^*;
\\
&&Y_{\nu\tilde N_i \tilde \nu^*}
= {1\over \sqrt{2}} (-g N_{i2}^* +\gp N_{i1}^*) ;
\\
&&Y_{t\tilde N_i \tilde t^*_j}
= L_{\tilde t_j}^* Y_{u\tilde N_i \tilde u^*_L} - 
R_{\tilde t_j}^* N_{i4}^* y_t;
\qquad\qquad
Y_{\overline t\tilde N_i \tilde t_j}
= R_{\tilde t_j} Y_{\overline u\tilde N_i \tilde u_R} 
- L_{\tilde t_j} N_{i4}^* y_t;
\\
&&Y_{b\tilde N_i \tilde b^*_j}
= L_{\tilde b_j}^* Y_{d\tilde N_i \tilde d^*_L} 
- R_{\tilde b_j}^* N_{i3}^* y_b;
\qquad\qquad
Y_{\overline b\tilde N_i \tilde b_j}
= R_{\tilde b_j} Y_{\overline d\tilde N_i \tilde d_R} 
- L_{\tilde b_j} N_{i3}^* y_b;
\\
&&Y_{\tau\tilde N_i \tilde \tau^*_j}
= L_{\tilde \tau_j}^* Y_{e\tilde N_i \tilde e^*_L} 
- R_{\tilde \tau_j}^* N_{i3}^* y_\tau;
\qquad\qquad
Y_{\overline \tau\tilde N_i \tilde \tau_j}
= R_{\tilde \tau_j} Y_{\overline e\tilde N_i \tilde e_R} 
-L_{\tilde \tau_j} N_{i3}^* y_\tau.
\phantom{xxxxx}
\eeq

The contributions from diagrams involving charginos, chiral fermions
and sfermions are:
\beq
V^{(2)}_{f \tilde C \tilde f'}
&=& \sum_{i=1}^2 \Biggl\lbrace
\sum_{j=1}^2 \biggl [
|Y_{\nu_\tau \tilde C_i \tilde \tau_j^*}|^2  \FFFS (0, \Ci, \stauj)
\nonumber \\ &&
+ 3 (|Y_{b\tilde C_i \tilde t_j^*}|^2 + |Y_{\overline b\tilde C_i \tilde
t_j}|^2) \FFFS (b, \Ci, \stopj)
+ 6 {\rm Re}[Y_{b\tilde C_i \tilde t_j^*}Y_{\overline
b\tilde C_i \tilde t_j}]
m_b m_{\tilde C_i} \FFFbS (b, \Ci, \stopj)
\nonumber \\ &&
+ 3 (|Y_{t\tilde C_i \tilde b_j^*}|^2 + |Y_{\overline t\tilde C_i \tilde
b_j}|^2) \FFFS (t, \Ci, \sbotj)
+ 6 {\rm Re}[Y_{t\tilde C_i \tilde b_j^*}
Y_{\overline t\tilde C_i \tilde b_j}]
m_t m_{\tilde C_i} \FFFbS (t, \Ci, \sbotj)
\biggr ]
\nonumber \\ &&
+ (|Y_{\tau\tilde C_i \tilde \nu_\tau^*}|^2 + |Y_{\overline \tau\tilde C_i
\tilde \nu_\tau}|^2) \FFFS (\tau, \Ci, \snutau )
+ 2 {\rm Re}[Y_{\tau\tilde C_i \tilde \nu_\tau^*}
Y_{\overline \tau \tilde C_i \tilde \nu_\tau}]
m_\tau m_{\tilde C_i} \FFFbS (\tau, \Ci, \snutau )
\nonumber \\ &&
+3 |Y_{d\tilde C_i \tilde u_L^*}|^2
[\FFFS (0, \Ci, \suL) + \FFFS (0, \Ci, \scL)]
\nonumber \\ &&
+ 3 |Y_{u\tilde C_i \tilde d_L^*}|^2
[\FFFS (0, \Ci, \sdL) + \FFFS (0, \Ci, \ssL)] 
\nonumber \\ &&
+ |Y_{e\tilde C_i \tilde \nu_e^*}|^2 
[\FFFS (0, \Ci, \snue) + \FFFS (0, \Ci, \snumu)]
\nonumber \\ &&
+ |Y_{\nu_e\tilde C_i \tilde e_L^*}|^2 
[\FFFS (0, \Ci, \seL) + \FFFS (0, \Ci, \smuL)]
\Biggr\rbrace .
\label{VfsfC}
\eeq
Here, the non-zero fermion-chargino-sfermion couplings are:
\beq
&&Y_{d\tilde C_i \tilde u_L^*} = Y_{e\tilde C_i \tilde \nu_e^*} 
= Y_{\tau\tilde C_i \tilde \nu_\tau^*} = -g V_{i1}^*;
\qquad\qquad 
Y_{u\tilde C_i \tilde d_L^*} = Y_{\nu_e \tilde C_i \tilde e_L^*} 
= -g U_{i1}^*;\phantom{xxxxx}
\\
&&Y_{b\tilde C_i \tilde t_j^*} = -L^*_{\tilde t_j} g V_{i1}^* 
+ R^*_{\tilde t_j} V_{i2}^* y_t ;
\qquad\qquad\>\>\>\>\>\>\>
Y_{\overline b\tilde C_i \tilde t_j} = L_{\tilde t_j} U_{i2}^* y_b;
\\
&&Y_{t\tilde C_i \tilde b_j^*} = -L^*_{\tilde b_j} g U_{i1}^* 
+ R^*_{\tilde b_j} U_{i2}^* y_b;
\qquad\qquad\>\>\>\>\>\,
Y_{\overline t\tilde C_i \tilde b_j} = L_{\tilde b_j} V_{i2}^* y_t;
\\
&&Y_{\nu_\tau \tilde C_i \tilde \tau_j^*} =
-L^*_{\tilde \tau_j} g U_{i1}^* + R^*_{\tilde \tau_j} U_{i2}^* y_\tau;
\qquad\qquad\>\>\>\>
Y_{\overline \tau\tilde C_i \tilde \nu_\tau} = y_\tau U_{i2}^* .
\eeq

The contributions from diagrams involving Higgs scalars and Standard Model
fermions are:
\beq
V^{(2)}_{ff\phi^0}
&=&
{1\over 2} \sum_{\phi^0}
\biggl \lbrace  
{3y_t^2} [|k_{u\phi^0}|^2 \FFFS (t,t,\phi^0) 
+ (k_{u\phi^0})^2 m_t^2 \FFFbS (t,t,\phi^0)]
\nonumber \\ &&
+ {3y_b^2} [|k_{d\phi^0}|^2 \FFFS (b,b,\phi^0) 
+ (k_{d\phi^0})^2 m_b^2 \FFFbS (b,b,\phi^0)]
\nonumber \\ &&
+ {y_\tau^2} [|k_{d\phi^0}|^2 \FFFS (\tau,\tau,\phi^0) 
+ (k_{d\phi^0})^2 m_\tau^2 \FFFbS (\tau,\tau,\phi^0)]
\biggr\rbrace;
\label{Vffphizero}
\\
V^{(2)}_{ff'\phi^\pm} &=& \sum_{\phi^\pm} \biggl \lbrace
3 (y_t^2 k^2_{u\phi^+} + y_b^2 k^2_{d\phi^+}) \FFFS (t,b,\phi^\pm) 
+ 6 k_{u\phi^+} k_{d\phi^+} y_t y_b m_t m_b \FFFbS (t,b,\phi^\pm)
\nonumber \\ &&
+ y_\tau^2 k^2_{d\phi^+} \FFFS (0,\tau,\phi^\pm) \biggr\rbrace .
\label{Vffphipm}
\eeq

The contributions from diagrams which involve a Higgs scalar propagator 
and two chargino and/or neutralino propagators are:
\beq
V^{(2)}_{\tilde C \tilde C \phi^0} &=& \sum_{i,j=1}^2 \sum_{\phi^0}
\biggl \lbrace
|Y_{\tilde C_i^+ \tilde C_j^- \phi^0}|^2 \FFFS (\Ci,\Cj,\phi^0)
\nonumber \\ && 
+ {\rm Re}[Y_{\tilde C_i^+ \tilde C_j^-\phi^0}
           Y_{\tilde C_j^+ \tilde C_i^- \phi^0}] 
m_{\tilde C_i} m_{\tilde C_j} \FFFbS (\Ci,\Cj,\phi^0)\biggr\rbrace;
\phantom{xx} 
\\
V^{(2)}_{\tilde N \tilde N \phi^0} &=&
{1\over 2}\sum_{i,j=1}^4 \sum_{\phi^0}\biggl \lbrace 
|Y_{\tilde N_i \tilde N_j \phi^0}|^2 \FFFS(\Ni,\Nj,\phi^0) 
\nonumber \\ && 
+ {\rm Re}[(Y_{\tilde N_i \tilde N_j \phi^0})^2] 
m_{\tilde N_i} m_{\tilde N_j} \FFFbS(\Ni,\Nj,\phi^0) \biggr \rbrace;
\phantom{xxx}
\\
V^{(2)}_{\tilde C \tilde N \phi^\pm} &=& 
\sum_{i=1}^2 \sum_{j=1}^4 \sum_{\phi^\pm} \biggl \lbrace
(|Y_{\tilde C_i^+ \tilde N_j \phi^-}|^2 
+|Y_{\tilde C_i^- \tilde N_j \phi^+}|^2) \FFFS (\Ci, \Nj, \phi^\pm)
\nonumber \\ &&
+ 2 {\rm Re}[Y_{\tilde C_i^+ \tilde N_j \phi^-}
Y_{\tilde C_i^- \tilde N_j \phi^+}]
m_{\tilde C_i} m_{\tilde N_j} \FFFbS (\Ci, \Nj, \phi^\pm) \biggr\rbrace,
\phantom{xxx}
\eeq
where the necessary couplings are:
\beq
Y_{\tilde C_i^+ \tilde C_j^- \phi^0} &=&
-{g\over \sqrt{2}}[k_{d\phi^0}^* V_{i1}^* U_{j2}^* 
                 + k_{u\phi^0}^* V_{i2}^* U_{j1}^*];
\\
Y_{\tilde N_i \tilde N_j \phi^0} &=&
{1\over 2}(g N_{i2}^* - \gp N_{i1}^*)
(k^*_{u\phi^0} N_{j4}^* - k^*_{d\phi^0} N_{j3}^*) 
+ (i \leftrightarrow j);
\\
Y_{\tilde C_i^+ \tilde N_j \phi^-} &=&
k_{u\phi^+} \Bigl [ g V_{i1}^* N_{j4}^* 
+ {1\over \sqrt{2}} V_{i2}^* (g N_{j2}^* + \gp N_{j1}^*)
\Bigr ] ;
\\
Y_{\tilde C_i^- \tilde N_j \phi^+} &=&
k_{d\phi^+} \Bigl [g U_{i1}^* N_{j3}^* 
-{1\over \sqrt{2}} U_{i2}^* (g N_{j2}^* + \gp N_{j1}^*)
\Bigr ] .
\eeq

\subsection{$SSV$-diagram contributions}
\label{subsec:SSV}

In this subsection I list the contributions to the MSSM two-loop effective
potential coming from diagrams involving one vector and two scalar
propagators. These all involve the function $\FSSV(x,y,z)$.

The contributions from diagrams involving the gluon and squarks are:
\beq
V^{(2)}_{\tilde q\tilde q g} &=& 2 g_3^2 \sum_{\tilde q}
\FSSV(\stilde q, \stilde q,0) .
\eeq

The contributions from diagrams involving the photon and sfermions are:
\beq
V^{(2)}_{\tilde f\tilde f \gamma} &=& 
{g^2 \gp^2 \over 2(g^2 + \gp^2)} \sum_{\tilde f} n_{\tilde f}
Q^2_{\tilde f} \FSSV(\stilde f, \stilde f,0) .
\eeq
Here $Q_{\tilde f}$ denotes the electric charge of the sfermion.

The contributions from diagrams involving the $W,Z$ bosons and
sfermions are:
\beq
V^{(2)}_{\tilde f\tilde f Z} &=&
{1\over 24 (g^2 + \gp^2)} \Biggl \lbrace
(3 g^2 - \gp^2)^2 [\FSSV (\suL, \suL, Z)+\FSSV (\scL, \scL, Z)]
\nonumber \\ &&
+ (3 g^2 + \gp^2)^2 [\FSSV (\sdL, \sdL, Z) + \FSSV (\ssL, \ssL, Z)]
\nonumber \\ &&
+ 16 \gp^4 [\FSSV (\suR, \suR, Z)+\FSSV (\scR, \scR, Z)]
\nonumber \\ &&
+ 4 \gp^4 [\FSSV (\sdR,\sdR,Z) + \FSSV (\ssR,\ssR,Z)]
\nonumber \\ &&
+ 3 (g^2 - \gp^2 )^2 [\FSSV (\seL,\seL,Z) + \FSSV (\smuL,\smuL,Z)]
\nonumber \\ &&
+ 3 (g^2 + \gp^2)^2 [\FSSV (\snue,\snue,Z)
+ \FSSV (\snumu,\snumu,Z) + \FSSV (\snutau,\snutau,Z) ]
\nonumber \\ &&
+ 12 \gp^4 [\FSSV(\seR,\seR,Z) + \FSSV(\smuR,\smuR,Z)] 
\nonumber \\ &&
+ \sum_{i,j=1}^2 |(3 g^2 -\gp^2) L_{\tilde t_i} L_{\tilde t_j}^*-4 \gp^2 R_{\tilde t_i} R_{\tilde t_j}^*|^2
\FSSV (\stopi,\stopj,Z)
\nonumber \\ &&
+ \sum_{i,j=1}^2 |(3 g^2 +\gp^2) L_{\tilde b_i} L_{\tilde b_j}^*-2 \gp^2 R_{\tilde b_i} R_{\tilde b_j}^*|^2
\FSSV (\sboti,\sbotj,Z)
\nonumber \\ &&
+ 3 \sum_{i,j=1}^2 |(g^2 -\gp^2) L_{\tilde \tau_i} L_{\tilde \tau_j}^*-2 \gp^2 R_{\tilde \tau_i} R_{\tilde
\tau_j}^*|^2 \FSSV (\staui,\stauj,Z)
\Biggr \rbrace;
\\
V^{(2)}_{\tilde f\tilde f' W} &=&
{g^2 \over 2} \biggl [
3 \FSSV (\suL,\sdL,W) 
+ 3 \FSSV (\scL,\ssL,W) 
+ 3 \sum_{i,j=1}^2 |L_{\tilde t_i} L_{\tilde b_j}|^2 \FSSV (\stopi,\sbotj,W)
\nonumber \\ &&
+ \FSSV (\snue,\seL,W) 
+ \FSSV (\snumu,\smuL,W) 
+ \sum_{i=1}^2 |L_{\tilde \tau_i}|^2 \FSSV (\snutau, \staui,W)
\biggr ] .
\eeq

The contributions from diagrams involving Higgs scalars and electroweak
gauge bosons are:
\beq
V^{(2)}_{\phi^\pm\phi^\pm \gamma} &=&
{g^2 \gp^2 \over 2 (g^2 + \gp^2)}\biggl [
\FSSV (\Hpm,\Hpm,0) + 
\FSSV (\Gpm,\Gpm,0) \biggr ];
\\
V^{(2)}_{\phi^\pm\phi^\pm Z} &=& {(g^2 -\gp^2)^2 \over 8 (g^2 + 
\gp^2)}\biggl [
\FSSV (\Hpm,\Hpm,Z) + 
\FSSV (\Gpm,\Gpm,Z) \biggr ];
\\
V^{(2)}_{\phi^0\phi^0 Z} &=&{g^2 + \gp^2 \over 8} \biggl \lbrace 
( \calpha \cbetaO + \salpha \sbetaO)^2
\left [ \FSSV (h^0, A^0,Z) + \FSSV (H^0,G^0,Z) \right ]
\nonumber \\ &&
+ (\salpha \cbetaO - \calpha \sbetaO)^2
\left [ \FSSV (h^0, G^0,Z) + \FSSV (H^0,A^0,Z) \right ]
\biggr \rbrace;
\\
V^{(2)}_{\phi^0\phi^\pm W} &=&{g^2 \over 4} \biggl \lbrace
( \calpha \cbetapm + \salpha \sbetapm)^2
\left [ \FSSV (h^0, \Hpm,W) + \FSSV (H^0,\Gpm,W) \right ]
\nonumber \\ &&
+ ( \salpha \cbetapm - \calpha \sbetapm)^2
\left [ \FSSV (h^0, \Gpm,W) + \FSSV (H^0,\Hpm,W) \right ]
\nonumber \\ &&
+ ( \cbetaO \cbetapm + \sbetaO \sbetapm)^2
\left [ \FSSV (A^0, \Hpm,W) + \FSSV (G^0,\Gpm,W) \right ]
\nonumber \\ &&
+ ( \sbetaO \cbetapm - \cbetaO \sbetapm)^2
\left [ \FSSV (A^0, \Gpm,W) + \FSSV (G^0,\Hpm,W) \right ]
\biggr \rbrace .
\eeq

\subsection{$VS$-diagram contributions}
\label{subsec:VS}

In this subsection, I list the contributions coming from diagrams
with one vector and one scalar propagator. These all involve the
function $\FVS(x,z)$. This function vanishes when the 
vector squared mass variable $x$ is zero, so only the $W$ and $Z$ bosons
contribute.

The contributions from diagrams with one electroweak gauge boson
and one Higgs scalar propagator are:
\beq
V^{(2)}_{W\phi} &=&
{g^2 \over 4} \sum_{\phi^0} \FVS (W,\phi^0) +
{g^2 \over 2} \sum_{\phi^\pm} \FVS (W,\phi^\pm) ; 
\\
V^{(2)}_{Z\phi} &=&
{g^2 + \gp^2 \over 8} \sum_{\phi^0} \FVS(Z,\phi^0)
+ {(g^2 - \gp^2)^2 \over 4(g^2 + \gp^2 )} \sum_{\phi^\pm} \FVS (Z,\phi^\pm)
.
\eeq

The contributions from diagrams with one electroweak gauge boson and
one sfermion propagator are:
\beq
V_{W\stilde f}^{(2)} &=& g^2 \sum_{\tilde f}
n_{\tilde f} |x_{\tilde f}| \FVS(W, \tilde f) ;
\\
V_{Z \stilde f}^{(2)} &=& {1\over 12 (g^2 + \gp^2)} 
\Biggl \lbrace
16 \gp^4 
\Bigl [\FVS (Z,\suR ) + \FVS (Z,\scR ) 
+ |R_{\tilde t_1}|^2 \FVS (Z,\stopI) + |R_{\tilde t_2}|^2 \FVS (Z,\stopII)
\Bigr ]
\nonumber \\ &&
+ (3 g^2 - \gp^2 )^2 
\Bigl [\FVS (Z,\suL ) + \FVS (Z,\scL ) + 
|L_{\tilde t_1}|^2 \FVS (Z,\stopI) + |L_{\tilde t_2}|^2 \FVS (Z,\stopII)
\Bigr ]
\nonumber \\ &&
+ 4 \gp^4 
\Bigl [\FVS (Z,\sdR ) + \FVS (Z,\ssR ) 
+ |R_{\tilde b_1}|^2  \FVS (Z,\sbotI)+ |R_{\tilde b_2}|^2  \FVS (Z,\sbotII)
\Bigr ]
\nonumber \\ &&
+ (3 g^2 + \gp^2 )^2 
\Bigl [\FVS (Z,\sdL ) + \FVS (Z,\ssL )
+ |L_{\tilde b_1}|^2  \FVS (Z,\sbotI)+ |L_{\tilde b_2}|^2  \FVS (Z,\sbotII)
\Bigr ]
\nonumber \\ &&
+ 12 \gp^4 
\Bigl [\FVS (Z, \seR) +\FVS (Z, \smuR) 
+ |R_{\tilde \tau_1}|^2  \FVS (Z,\stauI)+ |R_{\tilde \tau_2}|^2  
\FVS (Z,\stauII)
\Bigr ] 
\nonumber \\ &&
+ 3 (g^2 - \gp^2)^2 
\Bigl [\FVS (Z,\seL )+\FVS (Z, \smuL)
+ |L_{\tilde \tau_1}|^2  \FVS (Z,\stauI)+ |L_{\tilde \tau_2}|^2  
\FVS (Z,\stauII)
\Bigr ]
\nonumber \\ &&
+ 3 (g^2 + \gp^2)^2 
\Bigl [\FVS (Z,\snue )+\FVS (Z, \snumu) + \FVS (Z,\snutau )
\Bigr ]
\Biggr \rbrace . \phantom{x}
\eeq

\subsection{$VVS$-diagram contributions}
\label{subsec:VVS}

The contributions from diagrams with one scalar and two vector
propagators are:
\beq
V_{VVS}^{(2)} &=& 
{g^2 \gp^2 \over 2 (g^2 + \gp^2)} \left \lbrace
(\cbetapm \vd + \sbetapm
\vu)^2
\left [ g^2 \FVVS(W,0,\Gpm) + \gp^2 \FVVS(W,Z,\Gpm) \right ]
\right.
\nonumber \\
&& \left.
+ (\cbetapm \vu - \sbetapm \vd)^2
\left [ g^2 \FVVS(W,0,\Hpm) + \gp^2 \FVVS(W,Z,\Hpm) \right ] \right
\rbrace
\phantom{xxx}
\nonumber \\ 
&& + 
{1 \over 8} (\calpha \vu - \salpha \vd)^2  \left [
(g^2 + \gp^2)^2 \FVVS (Z,Z,h^0) + 2 g^4 \FVVS (W,W,h^0) \right ] 
\nonumber \\
&& +
{1\over 8 } (\calpha \vd + \salpha \vu)^2 \left [
(g^2 + \gp^2)^2 \FVVS (Z,Z,H^0) + 2 g^4 \FVVS (W,W,H^0) \right ] .
\eeq

\subsection{$FFV$-diagram and $\FFbV$-diagram contributions}
\label{subsec:FFV}

In this section I list the contributions from diagrams with 
one vector boson and two fermion
propagators. These include cases with no chirality-flipping mass insertion
[involving the function $\FFFV(x,y,z)$] and those with two such mass 
insertions on different propagators [involving the function 
$\FFFbV(x,y,z)$].

The contributions from diagrams with two Standard Model fermions and one 
vector boson are:
\beq
V^{(2)}_{qqg} &=& 4 g_3^2 \left [ 
\FFFV (t,t,0) - m_t^2 \FFFbV (t,t,0)
+ \FFFV (b,b,0) - m_b^2 \FFFbV (b,b,0) \right ];
\\
V^{(2)}_{ff\gamma} &=& {g^2 \gp^2 \over g^2 + \gp^2} 
\biggl \lbrace 
{4\over 3} [\FFFV (t,t,0)- m_t^2 \FFFbV (t,t,0) ] 
\nonumber \\ &&
+ {1\over 3}[\FFFV (b,b,0) - m_b^2 \FFFbV (b,b,0)]
+ \FFFV (\tau,\tau,0) - m_\tau^2 \FFFbV (\tau,\tau,0)\biggr \rbrace ;
\phantom{xxx}
\\
V^{(2)}_{ffZ} &=& {1\over 24 (g^2 + \gp^2)} \biggl [
(51 g^4 + 6 g^2 \gp^2 + 83 \gp^4) \FFFV (0,0,Z)
\nonumber \\  &&
+ (9 g^4 - 6 g^2 \gp^2 + 17 \gp^4 ) \FFFV (t,t,Z)
+ 8 \gp^2 (3g^2 - \gp^2) m_t^2 \FFFbV (t,t,Z) 
\nonumber \\  &&
+ (9 g^4 + 6 g^2 \gp^2 + 5 \gp^4 ) \FFFV (b,b,Z)
+4 \gp^2 (3g^2 + \gp^2) m_b^2 \FFFbV (b,b,Z) 
\nonumber \\  &&
+ (3 g^4 - 6 g^2 \gp^2 + 15 \gp^4) \FFFV (\tau,\tau,Z) 
+ 12 \gp^2 (g^2 - \gp^2) m_\tau^2 \FFFbV (\tau,\tau,Z) 
\biggr ] ; \phantom{xxx}
\\
V^{(2)}_{ff'W} &=& {g^2 \over 2} \Biggl [
3 \FFFV (t,b,W) + \FFFV (\tau,0,W) + 8 \FFFV (0,0,W)
\Biggr ] .
\eeq

The contributions from diagrams involving charginos and/or neutralinos
and electroweak vector bosons are:
\beq
V^{(2)}_{\tilde C \tilde C \gamma} &=&
{g^2 \gp^2 \over g^2 + \gp^2} \sum_{i=1}^2 \biggl [ \FFFV (\Ci,\Ci,0)
- m_{\Ci}^2 \FFFbV (\Ci,\Ci,0) \biggr ] ;
\\
V^{(2)}_{\tilde N \tilde N Z} &=& 
{g^2 + \gp^2 \over 2}\sum_{i,j=1}^4
\biggl [ |O^{''L}_{ij}|^2 \FFFV (\Ni,\Nj,Z)
+ (O^{''L}_{ij})^2 m_{\Ni} m_{\Nj} \FFFbV (\Ni,\Nj,Z) \biggr ]
;
\\
V^{(2)}_{\tilde C\tilde C Z} &=&
{g^2 + \gp^2 \over 2} \sum_{i,j=1}^2
\biggl [
(|O^{'L}_{ij}|^2 + |O^{'R}_{ij}|^2) \FFFV (\Ci,\Cj,Z)
\nonumber \\ &&
- 2 O^{'L}_{ij} O^{'R*}_{ij} m_{\Ci} m_{\Cj} \FFFbV (\Ci,\Cj,Z)
\biggr ] 
;
\\
V^{(2)}_{\tilde N \tilde C W} &=&
g^2 \sum_{i=1}^4 \sum_{j=1}^2 \biggl \lbrace 
(|O^{L}_{ij}|^2 + |O^{R}_{ij}|^2) \FFFV (\Ni,\Cj,W)
\nonumber \\ &&
- 2 {\rm Re}[O^{L}_{ij} O^{R*}_{ij}] m_{\Ni} m_{\Cj} \FFFbV (\Ni,\Cj,W)
\biggr \rbrace
.
\eeq
where the necessary couplings are:
\beq
O^{''L}_{ij} &=& -{1\over 2} N_{i3} N_{j3}^* + {1\over 2} N_{i4} N_{j4}^*
; \phantom{xxx}
\\ 
O^{'L}_{ij} &=& -V_{i1} V_{j1}^* - {1\over 2} V_{i2} V_{j2}^* +
{\gp^2 \over g^2 + \gp^2 }\delta_{ij};
\\ 
O^{'R}_{ij} &=& -U_{i1}^* U_{j1} - {1\over 2} U_{i2}^* U_{j2} +
{\gp^2 \over g^2 + \gp^2 }\delta_{ij};
\\ 
O^{L}_{ij} &=& N_{i2} V_{j1}^* - {1\over \sqrt{2}} N_{i4} V_{j2}^* 
;
\\
O^{R}_{ij} &=& N^*_{i2} U_{j1} + {1\over \sqrt{2}} N^*_{i3} U_{j2} 
.
\eeq

There is also a contribution involving gluon and gluino propagators:
\beq
V^{(2)}_{\tilde g \tilde g g} = 12 g_3^2 \left [ 
\FFFV (\stilde g, \stilde g,0) - |M_3|^2 \FFFbV (\stilde g, \stilde g,0)
\right ].
\eeq
This is independent of the VEVs $v_u$ and $v_d$, and therefore is 
irrelevant to the minimization of the effective potential. However,
it is $Q$-dependent, and therefore must be included
when renormalization group consistency is checked.

\subsection{Pure gauge contributions}
\label{subsec:gauge}

The contributions involving only vector and ghost fields are
\beq
V_{\rm gauge}^{(2)} = {g^2 \over 2 (g^2 + \gp^2)} \left [
\gp^2 \Fgauge (W,W,0) + g^2 \Fgauge (W,W,Z) \right ].
\eeq

This concludes the list of contributions to the two-loop effective
potential in the MSSM.  Partial results for the two-loop contributions in
the approximation that $g$, $\gp$, $y_\tau$, and $a_\tau$ vanish and there
are no CP-violating phases had previously been given in
\cite{Zhang:1999bm}-\cite{Espinosa:2000df}. (Applications of these results
to the Higgs scalar boson mass spectrum have been made in
refs.~\cite{Espinosa:2001mm}-\cite{Frank:2002qf}.) The two-loop
contributions to the effective potential involving only Standard Model
fields were given in \cite{Ford:1992pn}, but in the \MSbar scheme,
which is not convenient for the supersymmetric extension.

\section{Supersymmetric limits}
\label{sec:susylimits}
\setcounter{equation}{0}
\setcounter{footnote}{1}

The results of section \ref{sec:potential} can be checked by
considering non-realistic limits in which supersymmetry is restored.
Unbroken global supersymmetry requires that the effective potential 
vanishes. 

One such limit occurs if all supersymmetry breaking parameters are
0, and $v_u=v_d=0$. The only massive particles in the theory are then
the members of
the Higgs supermultiplets, with a common squared mass $x = |\mu|^2$. 
The one-loop effective potential is easily seen to vanish, since it
is a supertrace over the squared masses. From the results of section 
\ref{sec:potential}, one finds that:
\beq
V^{(2)} \! &=& \!{3 g^2 + \gp^2 \over 2} \Bigl [ 
\FSS (x,x) + 4 \FFFS (0,x,x) + \FSSV (x,x,0) + \FFFV (x,x,0) -
x \FFFbV (x,x,0) \Bigr ]
\nonumber \\ &&
+ 2 (3 y_t^2 + 3 y_b^2 + y_\tau^2) \Bigl [ 
x \FSSS(0,0,x) + \FFFS(0,0,x) + 2 \FFFS(0,x,0) \Bigr ] .
\eeq
Each of the quantities in brackets indeed vanishes, using the
expressions in ref.~\cite{general}.

Another way of maintaining supersymmetry is to again
make all supersymmetry-breaking parameters 0, but now also
require $\mu=0$, and take the VEVs along a $D$-flat direction $v_u=v_d$.
The parameters $g,\gp,g_3, y_t, y_b, y_\tau$ remain arbitrary.
Then one finds that $V^{(0)} = V^{(1)} = 0$. However, the two-loop
contribution from section \ref{sec:potential} does not vanish:
\beq
V^{(2)} = {g^2 + \gp^2 \over 8} \left [ 3J(m_t^2) - 3 J(m_b^2) - 
J(m_\tau^2) \right ]^2,
\eeq
where
\beq
J(x) = x \left [ {\rm ln}(x/Q^2) -1 \right ].
\eeq
This might seem to violate the lore that if supersymmetry is not broken
at tree-level, it remains unbroken in perturbation theory. The 
resolution is that the classical $D$-flat condition $v_u=v_d$ is 
perturbed, 
because of the isospin violation of the
Yukawa couplings. The true flat direction is 
parameterized by:
\beq
v_u^2 = v_d^2 + {1\over 16 \pi^2} \left [ 3J(m_t^2) - 3 J(m_b^2) -
J(m_\tau^2) \right ].
\eeq
One then finds that $V_{\rm eff}$ is indeed 0 up to terms of
three-loop order; $V^{(0)}$, $V^{(1)}$, $V^{(2)}$
contribute to $V_{\rm eff}$ in the ratio $1: -2:1$. 

\section{Scale dependence of the effective potential and running of 
$\Lambda$, $\vu$ and $\vd$}
\label{sec:scale}
\setcounter{equation}{0}
\setcounter{footnote}{1}

An important consistency check on the effective potential formalism is
that the value of $V_{\rm eff}$, being a physical quantity, should be 
independent of the arbitrary choice of renormalization scale $Q$. 
In the MSSM, the
equation which expresses this is:
\beq
Q{d \over dQ} V_{\rm eff} = \left [ Q {\partial \over \partial Q} +
\sum_{\lambda} \beta_\lambda {\partial \over \partial \lambda} 
-\gamma_{H_u}^{(S)} v_u {\partial \over \partial v_u}
-\gamma_{H_d}^{(S)} v_d {\partial \over \partial v_d}
\right ] V_{\rm eff} = 0.
\eeq
Here $\lambda$ represents all of the running \DRbarprime parameters of the
theory (except the VEVs), and $\gamma_{H_u}^{(S)}, \gamma_{H_d}^{(S)}$
are the anomalous dimensions of the Higgs scalar fields in Landau gauge.
Note that these anomalous dimensions are gauge-dependent, and differ from 
the anomalous dimensions of the chiral superfields, because of gauge
fixing. (See for example refs.~\cite{Jones:mz} and 
\cite{Yamada:2001ck}.) Using the general
formulas given in ref.~\cite{general}, I obtain:
\beq
\gamma_{H_u}^{(S)} &=& 
{1 \over 16 \pi^2} \gamma_{H_u}^{(S,1)} 
+ {1 \over (16 \pi^2)^2} \gamma_{H_u}^{(S,2)};
\\
\gamma_{H_d}^{(S)} &=& 
{1 \over 16 \pi^2} \gamma_{H_d}^{(S,1)} 
+ {1 \over (16 \pi^2)^2} \gamma_{H_d}^{(S,2)},
\eeq
where
\beq
\gamma_{H_u}^{(S,1)} &=& 3 {\rm Tr}[\Yu^\dagger \Yu] - {3 \over 4} g^2
- {1\over 4} \gp^2;
\\
\gamma_{H_u}^{(S,2)} &=& 
-9 {\rm Tr}[\Yu^\dagger \Yu \Yu^\dagger \Yu] 
-3 {\rm Tr}[\Yu^\dagger \Yu \Yd^\dagger \Yd]
+(16 g_3^2 + {4\over 3} \gp^2) {\rm Tr}[\Yu^\dagger \Yu] 
\nonumber \\ &&
+ 3 g^4 + {3\over 4} g^2 \gp^2 + {23\over 8} \gp^4 ;
\\
\gamma_{H_d}^{(S,1)} &=& 3 {\rm Tr}[\Yd^\dagger \Yd] 
+ {\rm Tr}[\Ye^\dagger \Ye] - {3 \over 4} g^2
- {1\over 4} \gp^2 ;
\\
\gamma_{H_d}^{(S,2)} &=& 
-9 {\rm Tr}[\Yd^\dagger \Yd \Yd^\dagger \Yd]
-3 {\rm Tr}[\Yu^\dagger \Yu \Yd^\dagger \Yd]
-3 {\rm Tr}[\Ye^\dagger \Ye \Ye^\dagger \Ye]
\nonumber \\ &&
+(16 g_3^2 - {2\over 3} \gp^2) {\rm Tr}[\Yd^\dagger \Yd]
+ 2 \gp^2 {\rm Tr}[\Ye^\dagger \Ye]
+ 3 g^4 + {3\over 4} g^2 \gp^2 + {23 \over 8} \gp^4.
\eeq
The \DRbarprime VEVs run with renormalization
scale according to:
\beq
Q{d v_u\over dQ}  &=& -\gamma_{H_u}^{(S)} v_u;
\label{vuRGE}
\\
Q{d v_d\over d Q} &=& -\gamma_{H_d}^{(S)} v_d.
\label{vdRGE}
\eeq

The \DRbarprime two-loop beta functions for all of the MSSM parameters can 
be found\footnote{The parameters $\Au$, $\Ad$, $\Ae$, and $b$ in the
present paper were represented by
the symbols
${\bf h}_u$, ${\bf h}_d$, ${\bf h}_e$ and $B$ 
in ref.~\cite{Martin:1994zk}. That reference also has an 
obvious overall minus sign error in the specification of the MSSM soft 
Lagrangian; ${\cal L}$ should be $-{\cal L}$ in eqs.~(4.2) and (4.3). The 
results given in  that 
paper are actually in the \DRbarprime scheme, although not explicitly 
identified as such at the time (see the ``Note added").}
in ref.~\cite{Martin:1994zk}. The exception is the renormalization group 
running of the
\DRbarprime field-independent vacuum energy, which can be obtained from 
ref.~\cite{general}:
\beq
\beta_{\Lambda} &=& {1\over 16 \pi^2} \beta_{\Lambda}^{(1)} 
+ {1\over (16 \pi^2)^2} \beta_{\Lambda}^{(2)}; \\
\beta_{\Lambda}^{(1)} &=& 
2 (m^2_{H_u})^2
+ 2 (m^2_{H_d})^2
+ 4 |\mu|^2 (m^2_{H_u} + m^2_{H_d})
+ 4 |b|^2 
\nonumber \\ && 
+ {\rm Tr}[6 \bm^2_Q \bm^2_Q +2 \bm^2_L \bm^2_L +3 \bm^2_u \bm^2_u + 3 
\bm^2_d \bm^2_d 
+ \bm^2_e \bm^2_e]
\nonumber \\ && 
- |M_1|^4 - 3 |M_2|^4 - 8 |M_3|^4 ;\\
\beta_{\Lambda}^{(2)} &=& 
44 \gp^2 |M_1|^4 + 36 g^2 |M_2|^4
+ 
|\mu|^2 
\left  
(24 g^2 |M_2|^2 + 8 \gp^2 |M_1|^2 \right ) 
\nonumber \\ && 
+ (6 g^2 + 2 \gp^2) \left \lbrace 2 |\mu|^2 (m^2_{H_u} + 
m^2_{H_d}) + (m^2_{H_u})^2 + (m^2_{H_d})^2 +
2 |b|^2 + {\rm Tr}[\bm_L^2 \bm_L^2] \right \rbrace
\nonumber \\ && 
+ (32 g_3^2 + 18 g^2 + {2 \over 3} \gp^2 ) {\rm Tr}[\bm_Q^2 \bm_Q^2] 
+ (16 g_3^2 + {16 \over 3} \gp^2 ) {\rm Tr}[\bm_u^2 \bm_u^2] 
\nonumber \\ && 
+ (16 g_3^2 + {4 \over 3} \gp^2 ) {\rm Tr}[\bm_d^2 \bm_d^2] 
+ 4 \gp^2  {\rm Tr}[\bm_e^2 \bm_e^2] 
\nonumber \\ && 
- (12 g^2 M_2 + 4 \gp^2 M_1) b^* \mu
- (12 g^2 M_2^* + 4 \gp^2 M_1^*) b \mu^*
\nonumber \\ && 
- 12 {\rm Tr}[\Yu^\dagger \Yu] \left \lbrace
(m^2_{H_u})^2 + |\mu|^2 (2 m^2_{H_u}  + m^2_{H_d}) + |b|^2
\right \rbrace
\nonumber \\ && 
- (
12 {\rm Tr}[\Yd^\dagger \Yd] + 4 {\rm Tr}[\Ye^\dagger \Ye])
\left \lbrace (m^2_{H_d})^2 +
|\mu|^2 (2 m^2_{H_d}  + m^2_{H_u})  + |b|^2 \right \rbrace 
\nonumber \\ && 
- 12 {\rm Tr}[ \bm^2_Q \bm^2_Q (\Yu^\dagger \Yu + \Yd^\dagger \Yd)]
- 12 {\rm Tr}[ \bm^2_u \bm^2_u \Yu \Yu^\dagger]
- 12 {\rm Tr}[ \bm^2_d \bm^2_d \Yd \Yd^\dagger]
\nonumber \\ && 
- 4 {\rm Tr}[ \bm^2_L \bm^2_L \Ye^\dagger \Ye]
- 4 {\rm Tr}[ \bm^2_e \bm^2_e \Ye \Ye^\dagger]
\nonumber \\ && 
- |\mu|^2 \Bigl \lbrace 12 {\rm Tr}[ \bm^2_Q (\Yu^\dagger \Yu + 
\Yd^\dagger \Yd)]
+ 12 {\rm Tr}[ \bm^2_u \Yu \Yu^\dagger]
\nonumber \\ && 
+ 12 {\rm Tr}[ \bm^2_d \Yd \Yd^\dagger]
+ 4 {\rm Tr}[ \bm^2_L  \Ye^\dagger \Ye]
+ 4 {\rm Tr}[ \bm^2_e \Ye \Ye^\dagger] \Bigr \rbrace 
\nonumber \\ && 
- 12 {\rm Tr}[ \bm^2_Q (\Au^\dagger \Au + \Ad^\dagger \Ad)]
- 12 {\rm Tr}[ \bm^2_u \Au \Au^\dagger]
- 12 {\rm Tr}[ \bm^2_d \Ad \Ad^\dagger]
\nonumber \\ && 
- 4 {\rm Tr}[ \bm^2_L  \Ae^\dagger \Ae]
- 4 {\rm Tr}[ \bm^2_e \Ae \Ae^\dagger]
- 12 {\rm Tr}[\Au^\dagger \Au] (m^2_{H_u} + |\mu|^2)
\nonumber \\ && 
- (12 {\rm Tr}[\Ad^\dagger \Ad] + 4 {\rm Tr}[\Ae^\dagger \Ae])
( m^2_{H_d} +|\mu|^2)
- 12 {\rm Tr}[\Yu^\dagger \Au] \mu b^*
- 12 {\rm Tr}[\Au^\dagger \Yu] \mu^* b
\nonumber \\ && 
- (12 {\rm Tr}[\Yd^\dagger \Ad] + 4 {\rm Tr}[\Ye^\dagger \Ae])
\mu b^*
- (12 {\rm Tr}[\Ad^\dagger \Yd] + 4 {\rm Tr}[\Ae^\dagger \Ye])
\mu^* b .
\eeq
Using these equations, I have checked that the effective potential
found in section \ref{sec:potential} is indeed renormalization 
scale-invariant at two-loop order. This demonstration is tedious
and omitted.

\section{A numerical example}
\label{sec:example}
\setcounter{equation}{0}
\setcounter{footnote}{1}

I now present a quasi-realistic numerical example, to 
illustrate the results above. As a template model, I choose parameters at
a renormalization scale $Q_0=640$ GeV:
\beq
&&
\gp = 0.36,\>\> g=0.65,\>\> g_3 = 1.06, \>\>
y_t = 0.90,\>\> y_b = 0.13,\>\> y_\tau = 0.10, 
\nonumber \\ && 
M_1=150,\>\> M_2 = 280,\>\> M_3 = 800,\>\>\> 
a_t = -600,\>\> a_b = -150,\>\> a_\tau=-40 \>\>\,\mbox{GeV}, 
\nonumber \\  && 
m^2_{Q_{1,2}} = (780)^2, \>\, 
m^2_{u_{1,2}} = (740)^2, \>\,
m^2_{d_{1,2}} = (735)^2, \>\,
m^2_{L_{1,2}} = (280)^2, \>\,
m^2_{e_{1,2}} = (200)^2\>\>\mbox{GeV}^2 ,
\phantom{xxx}
\nonumber \\ &&
m^2_{Q_3} = (700)^2,\>\> 
m^2_{u_3} = (580)^2,\>\> 
m^2_{d_3}= (725)^2,\>\> 
m^2_{L_3}= (270)^2,\>\> 
m^2_{e_3} = (195)^2\>\>\,\mbox{GeV}^2 ,
\nonumber \\ &&
m^2_{H_u}= -(500)^2,\>\>
m^2_{H_d} = (270)^2\>\>\,\mbox{GeV}^2
\label{templateparams}
\eeq
and
\beq
\mu = 504.1811202\>\,{\rm GeV};
\qquad\qquad
b = (184.2202586\>\,{\rm GeV})^2 .
\label{templatemub}
\eeq
The last two values are engineered so that the minimum of the full
2-loop effective potential found in section \ref{sec:potential} is:
\beq
v_u(Q_0) = 172\>\,{\rm GeV};
\qquad\>\>
\qquad
v_d(Q_0) = 17.2\>\,{\rm GeV}.
\label{templatevuvd}
\eeq

\begin{figure}[tp] 
\centerline{
\epsfxsize=3.5in\epsfbox{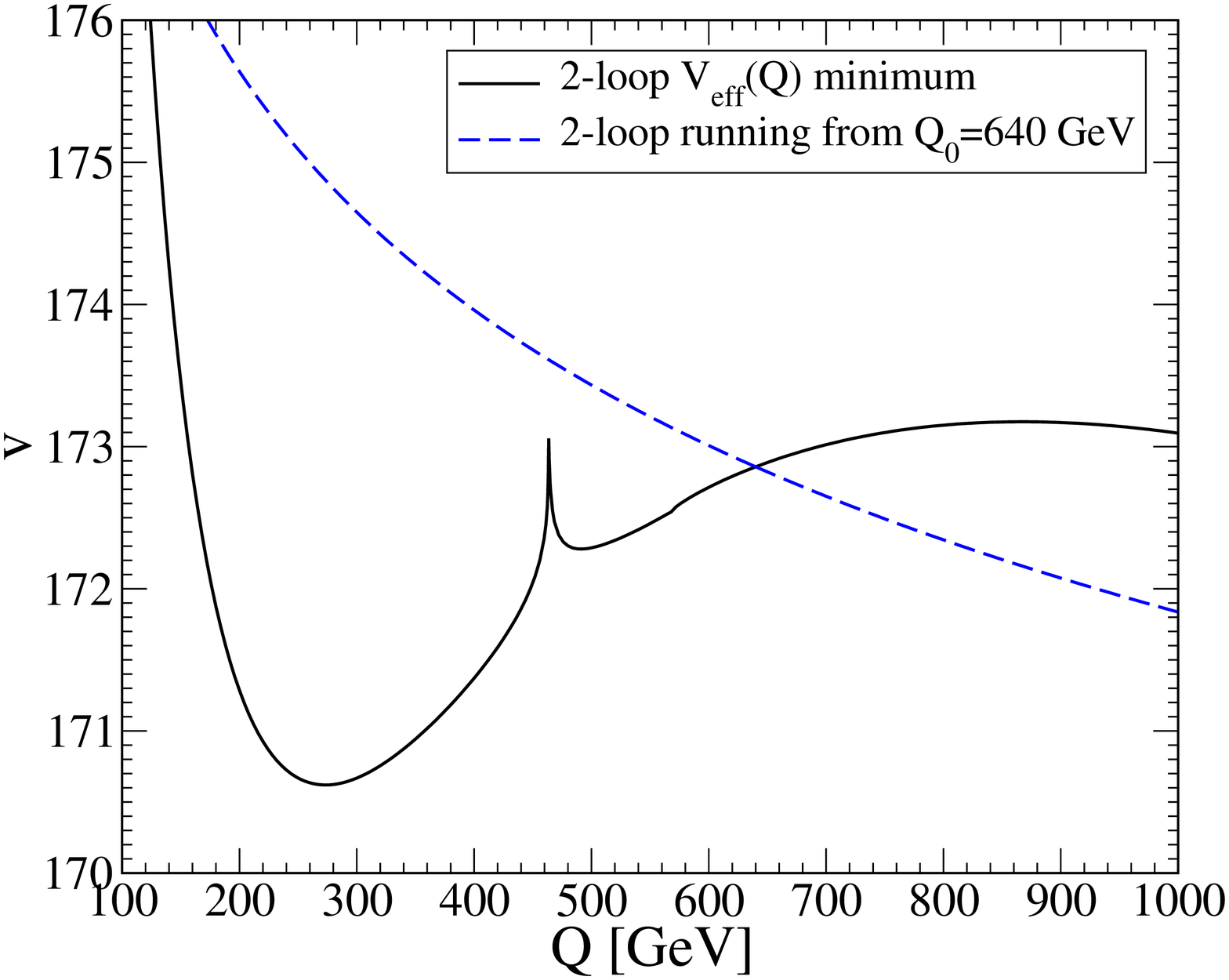} \hfil
\epsfxsize=3.5in\epsfbox{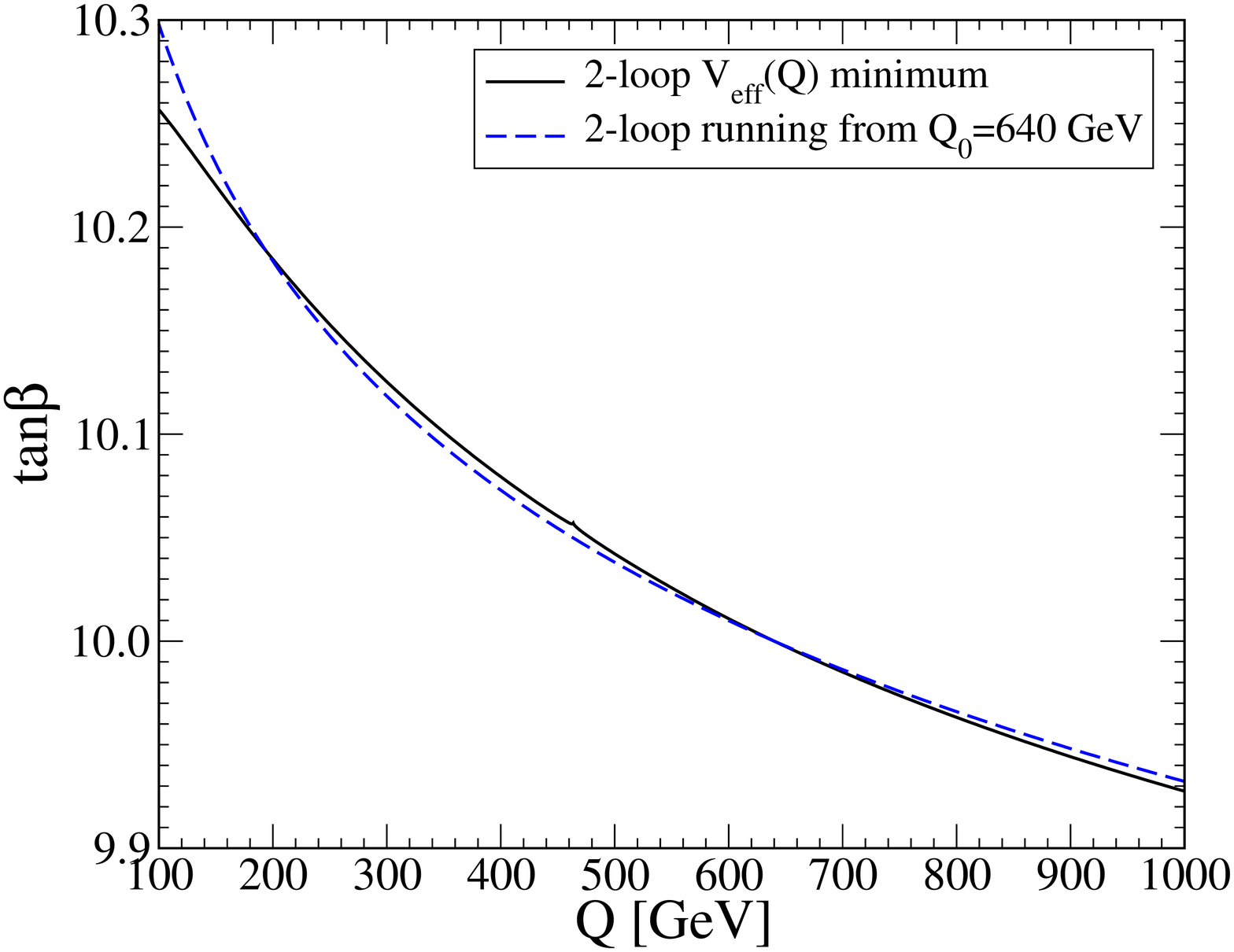}
}
\caption{Comparison of the values of $v(Q)$ and $\tan\beta(Q)$ as a test
of renormalization scale invariance. The solid lines are values 
obtained directly from minimization of the two-loop effective potential
at the scale $Q$, using Lagrangian parameters run from $Q_0 = 640$ GeV. 
The dashed lines are the result obtained by
renormalization group running of the template VEVs of 
eq.~(\ref{templatevuvd}) from $Q_0$ to $Q$ using 
eqs.~(\ref{vuRGE}),(\ref{vdRGE}). The discrepancy is due 
to effects at three loop order and beyond.}
\label{fig:vevrun}
\end{figure}

One way to test the accuracy of the effective potential is by checking
scale invariance. While I have done this analytically at two-loop order
as described in section \ref{sec:scale}, in practice the neglected effects 
of higher order can be quite significant. To study this, I 
run the Lagrangian parameters of 
eqs.~(\ref{templateparams}),(\ref{templatemub}) from the
template scale $Q_0$ to another scale $Q$, using the two-loop
renormalization group equations of ref.~\cite{Martin:1994zk}. The minimum 
of 
the 
effective potential at this new scale is found numerically.
The resulting VEVs are compared to the values obtained by running
the values of eq.~(\ref{templatevuvd}) for $v_u$ and $v_d$ obtained at 
$Q_0$ to $Q$ using 
eqs.~(\ref{vuRGE}),(\ref{vdRGE}).
The results of this comparison are shown in terms of the 
running quantities:
\beq
v(Q) \equiv \sqrt{v_u^2 + v_d^2};\qquad\qquad\>
\tan\beta(Q) \equiv  v_u/v_d
\eeq
in figure \ref{fig:vevrun}.
The good news is that the comparison of $v(Q)$ obtained by the two 
methods shows agreement to better than 0.5\% for a significant range
of scales $Q$ near the template scale $Q_0$ (which is approximately
$\sqrt{m_{\tilde t_1} m_{\tilde t_2}}$), and the comparison of
$\tan\beta$ is good to better than $0.1$\%. However, near $Q_0$ 
the slopes of
$v(Q)$ found by the two methods actually have the opposite sign!
This surprising sensitivity to higher-loop effects is a consequence
of the shallowness of the scalar potential along the direction
$v_u/v_d \approx \tan\beta$. In contrast, higher-loop contributions
have relatively much less effect on $v_u/v_d$, since that
corresponds to a much steeper direction of the potential.

An aside: the cusp-like feature found near $Q= 463$
GeV occurs because, at that scale, the tree-level squared mass of
$h^0$ at the minimum of the two-loop effective potential goes through 0; 
it is positive for 
all larger $Q$. This leads
to significant numerical effects because of the appearance of 
terms involving ln$(m_{h^0}^2)$ in the effective potential:
\beq
V^{(1)} &=& {1\over 4} (m_{h^0}^2)^2 {\rm ln}(m_{h^0}^2/Q^2) + \ldots;\\
V^{(2)} &=&  m_{h^0}^2 \left [
c_1 {\rm ln}(m_{h^0}^2/Q^2) + 
c_2 {\rm ln}^2(m_{h^0}^2/Q^2) 
\right ]
+ 
\ldots
.
\eeq
where $c_1$ and $c_2$ have dimensions of (mass)$^2$. Thus $V_{\rm eff}$ is
well-defined in the limit $m_{h^0}^2 \rightarrow 0$. However, while the
first derivatives of $V^{(1)}$ with respect to $v_u$ and $v_d$ are
finite, the first derivatives of $V^{(2)}$ have logarithmic and
double-logarithmic divergences in that limit.

\begin{figure}[tp]
\centerline{
\epsfxsize=3.5in\epsfbox{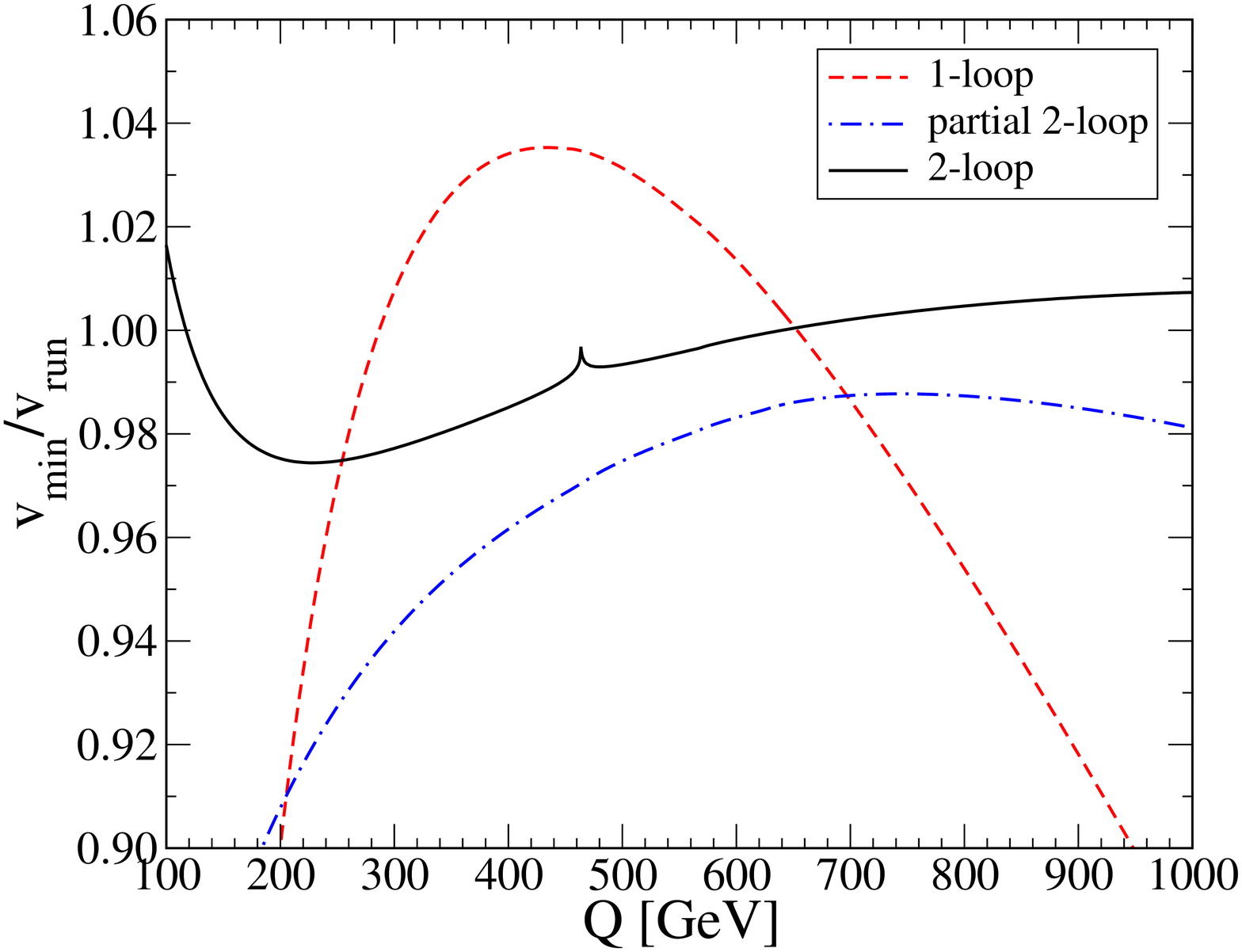} \hfil
\epsfxsize=3.5in\epsfbox{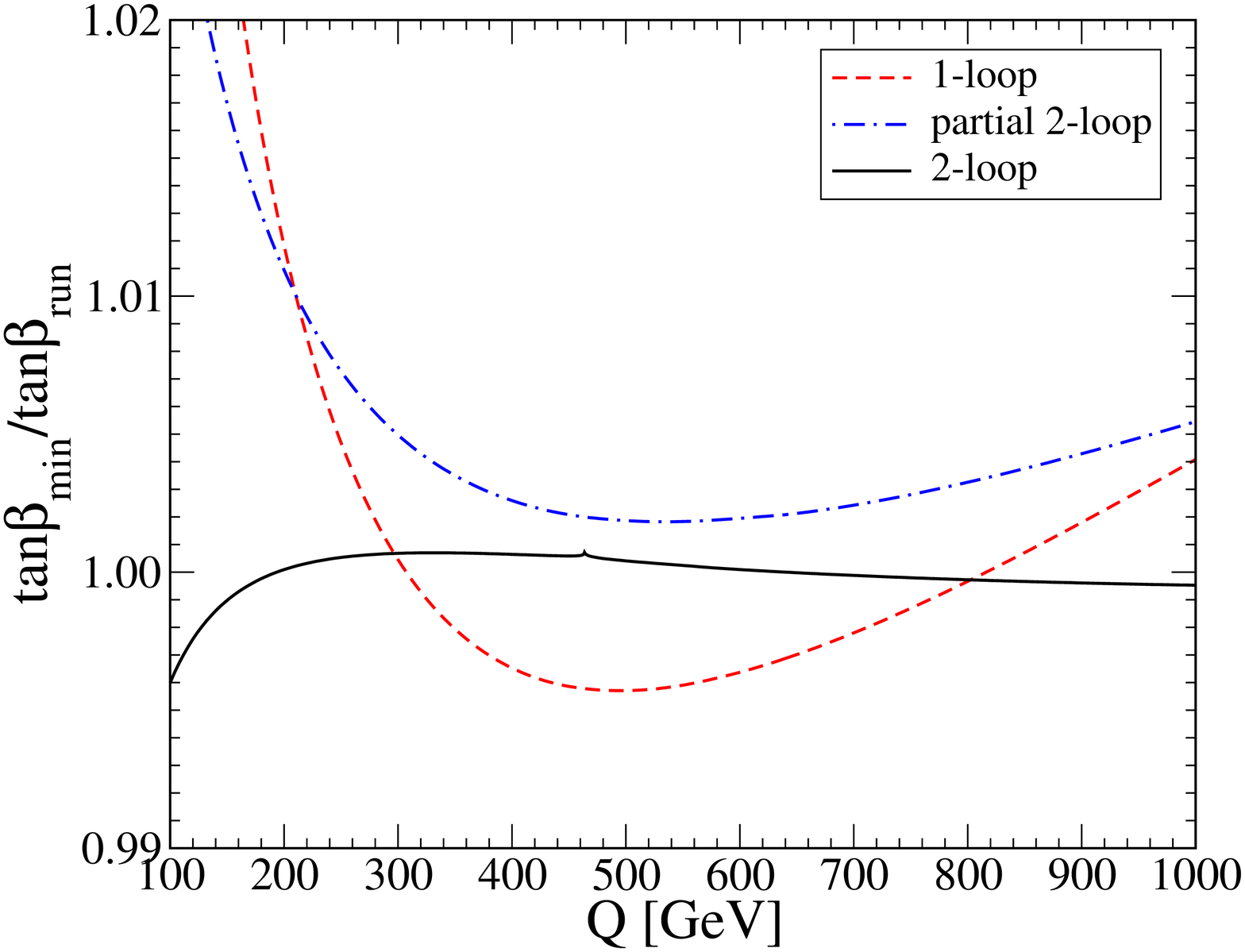}
}
\caption{Comparisons of the values and scale dependence of the VEV 
parameters $v$ and $\tan\beta$ found by
minimizing the effective potential in different approximations.
The label ``min" refers to values obtained by direct minimization at $Q$, 
and ``run" means obtained by running the template values of 
eq.~(\ref{templatevuvd}) from $Q_0$.
The solid lines follow from the full two-loop effective potential
found in section 3 of the present paper. The dashed lines represent the 
one-loop approximation, and the dot-dashed lines the partial two-loop 
approximation of refs.~\cite{Zhang:1999bm}-\cite{Espinosa:2000df}.}
\label{fig:vevratio}
\end{figure}

A comparison of the full two-loop effective potential to previous
approximations is shown in figure \ref{fig:vevratio}. The graphs show the 
ratios $v_{\rm min}/v_{\rm run}$ and 
$\tan\beta_{\rm min}/\tan\beta_{\rm run}$, where ``min"
means obtained by direct minimization at $Q$, and ``run" means 
obtained by running the template values of $v_u$, 
$v_d$ in eq.~(\ref{templatevuvd}) 
from $Q_0$ to $Q$.
This shows that the full two-loop effective potential indeed
mitigates the scale dependence, and changes the value found for 
$v$ compared to
the previous state-of-the-art approximation in 
refs.~\cite{Zhang:1999bm}-\cite{Espinosa:2000df}
by nearly $1.5$\% in this example.

\begin{figure}[tp]
\centerline{
\epsfxsize=3.5in\epsfbox{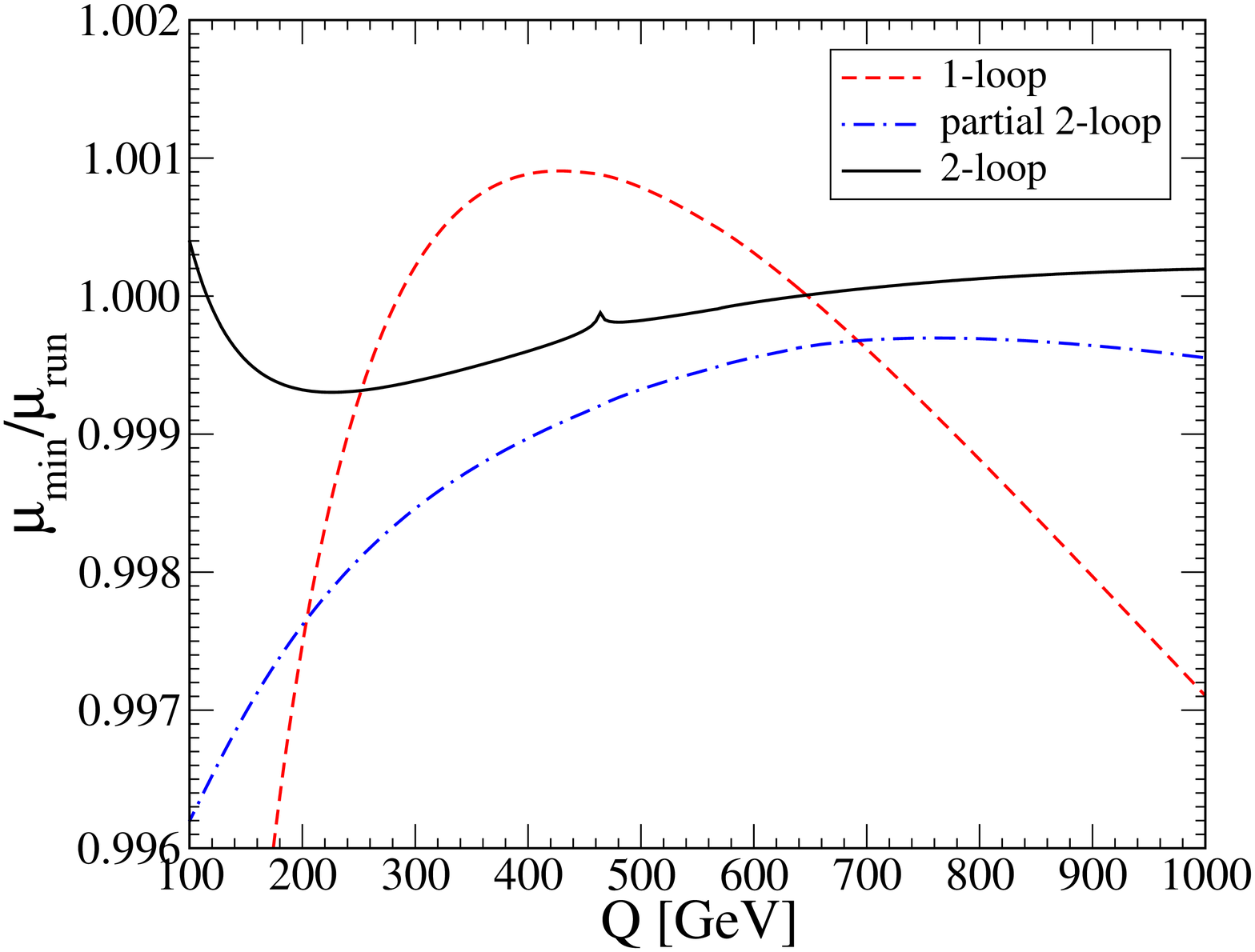} \hfil
\epsfxsize=3.5in\epsfbox{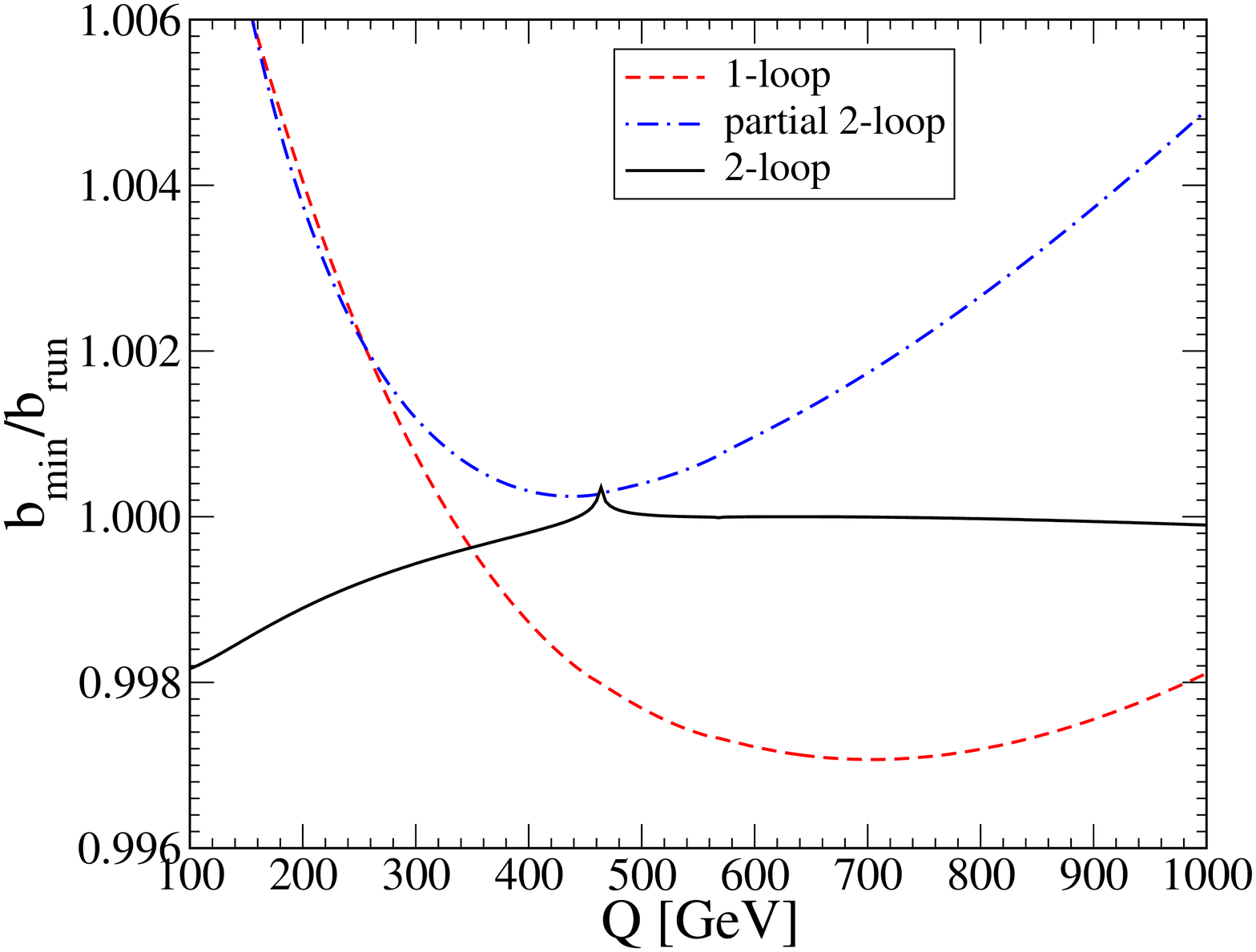}
}
\caption{Comparisons of the values and scale-dependence of the parameters
$\mu$ and $b$ obtained by requiring the effective 
potential to be minimized.
The label ``min" refers to values obtained by running the Lagrangian 
parameters of eq.~(\ref{templateparams}) and VEVs of 
eq.~(\ref{templatevuvd}) to $Q$, 
and ``run" means obtained by running the template values for $\mu$, $b$ of
eq.~(\ref{templatemub}) from $Q_0$ to $Q$.
The solid lines follow from the full two-loop effective potential
found in section 3. The dashed lines represent the one-loop approximation,
and the dot-dashed lines the partial two-loop approximation of
refs.~\cite{Zhang:1999bm}-\cite{Espinosa:2000df}.}
\label{fig:mubrun}
\end{figure}

A different way to look at things, which may be closer to the situation we
will face when confronted with experiment, is to view $v_u$ and
$v_d$ as input data rather than output parameters, and use
the two minimization conditions of the effective potential to extract 
values for two other parameters. Since $v_u$ and $v_d$
are especially sensitive to radiative corrections because of the 
shallowness of the
potential, treating them as among the known quantities is advantageous. 
Here I will follow the 
commonplace procedure of treating $b$ and $\mu$ as the unknowns
to be solved for, with given values of $v_u$, $v_d$ and all the other
Lagrangian parameters. (In reality, it seems clear that global fits to 
various observables will 
have to 
be conducted, since no more direct measurement of $m_{H_u}^2$ and 
$m_{H_d}^2$
is possible.) In figure \ref{fig:mubrun}, I have graphed the ratios of 
$\mu_{\rm min}/\mu_{\rm run}$ and
$b_{\rm min}/b_{\rm run}$, where ``min" means that the
effective potential at $Q$ is required to be minimized, using
values obtained by running the Lagrangian
parameters of eq.~(\ref{templateparams}) and VEVs of
eq.~(\ref{templatevuvd}) from $Q_0$ to $Q$,
while ``run" means those obtained by running the template 
values for $\mu$, $b$ of eq.~(\ref{templatemub}) from $Q_0$ to $Q$ using 
two-loop renormalization group equations.
In each case, the full two-loop results are compared to those obtained
using one-loop and partial two-loop approximations for the effective 
potential. In this model, the
scale-dependences of $\mu$ and $b$ are less than a few hundredths 
of a percent, using the full two-loop potential over a
wide range of scales $Q$.

\section{Outlook}
\label{sec:outlook}
\setcounter{equation}{0}
\setcounter{footnote}{1}

In this paper, I have presented the complete two-loop effective potential
for the MSSM in the Landau gauge and the \DRbarprime scheme.

The two-loop effective potential found here can, in principle, be
renormalization-group improved \cite{RGimprovement} to sum leading and
sub-leading logarithms of ratios of different mass scales. However, the
logarithmic contributions to the effective potential are typically not
overwhelming compared to the non-logarithmic ones. Renormalization group
improvement does give an improved scale dependence, because that is what
it is designed to do. However, improved scale-independence does not always
imply improved accuracy; it is a necessary but not sufficient criterion.
Therefore, the efficacy of further renormalization-group improvement of
the MSSM effective potential on top of the full two-loop result is unclear
to me.

If supersymmetry is part of the new physics associated with electroweak
symmetry breaking, then these results will be part of a program to fit
accurately experimental data to underlying Lagrangian parameters. Other
parts of this program will require more precise calculations of the
superpartner and Higgs scalar physical masses. Of particular importance is
an improved calculation of the physical $h^0$ mass, which is well-known to
be highly sensitive to radiative corrections. Unfortunately, $m_{h^0}^2$
cannot simply be obtained by taking the second derivatives of the
effective potential found here, because of important wave-function
renormalization effects.

The calculation described here can be extended to various non-minimal
versions of supersymmetry, for example those with additional singlet
fields. This can be done as a straightforward application of
ref.~\cite{general} for a general theory. The fact that $h^0$ was not
discovered at LEP may be taken as support for the importance of
considering such models. 

This work was supported in part by the NSF grant number PHY-9970691.


\end{document}